\documentclass[11pt,twoside]{article}
\usepackage{latexsym}
\makeatletter
									       
\def\icite{\@ifnextchar [{\@tempswatrue\@citey}{\@tempswafalse\@citey[]}}
\def\@citex[#1]#2{%
\if@filesw \immediate \write \@auxout {\string \citation {#2}}\fi 
\@tempcntb\m@ne \let\@h@ld\relax \def\@citea{}%
\@cite{%
  \@for \@citeb:=#2\do {%
    \@ifundefined {b@\@citeb}%
      {\@h@ld\@citea\@tempcntb\m@ne{\bf ?}%
      \@warning {Citation `\@citeb ' on page \thepage \space undefined}}%
      {\@tempcnta\@tempcntb \advance\@tempcnta\@ne%
      \@tempcntb\number\csname b@\@citeb \endcsname \relax%
      \ifnum\@tempcnta=\@tempcntb %   Number follows previous--hold on to it
	\ifx\@h@ld\relax%
	  \edef \@h@ld{\@citea\csname b@\@citeb\endcsname}% 
	\else%
	  \edef\@h@ld{\ifmmode{-}\else--\fi\csname b@\@citeb\endcsname}%
	\fi%
      \else%   %  non-successor--dump what's held and do this one
	\@h@ld\@citea\csname b@\@citeb \endcsname% 
	\let\@h@ld\relax%
      \fi}%
    \def\@citea{,\penalty\@highpenalty\,}%
  }\@h@ld
}{#1}}
\def\@citey[#1]#2{%
\if@filesw \immediate \write \@auxout {\string \citation {#2}}\fi 
\@tempcntb\m@ne \let\@h@ld\relax \def\@citea{}%
\@icite{%
  \@for \@citeb:=#2\do {%
    \@ifundefined {b@\@citeb}%
      {\@h@ld\@citea\@tempcntb\m@ne{\bf ?}%
      \@warning {Citation `\@citeb ' on page \thepage \space undefined}}%
      {\@tempcnta\@tempcntb \advance\@tempcnta\@ne%
      \@tempcntb\number\csname b@\@citeb \endcsname \relax%
      \ifnum\@tempcnta=\@tempcntb %   Number follows previous--hold on to it
	\ifx\@h@ld\relax%
	  \edef \@h@ld{\@citea\csname b@\@citeb\endcsname}% 
	\else%
	  \edef\@h@ld{\ifmmode{-}\else--\fi\csname b@\@citeb\endcsname}%
	\fi%
      \else%   %  non-successor--dump what's held and do this one
	\@h@ld\@citea\csname b@\@citeb \endcsname% 
	\let\@h@ld\relax%
      \fi}%
    \def\@citea{,\penalty\@highpenalty\,}%
  }\@h@ld
}{#1}}
\def\@cite#1#2{{$^{#1}$\if@tempswa , #2\fi }}
\def\@icite#1#2{{$#1$\if@tempswa , #2\fi }}

\gdef\@publabel{\hfil}
\gdef\@pubdate{\null}
\gdef\@pubnumber{\null}
\gdef\@author{\null}
\gdef\@title{\null}
\gdef\@abstract{\null}
\long\def\pubdate#1{\gdef\@pubdate{#1}}
\long\def\pubnumber#1{\gdef\@pubnumber{#1}}
\long\def\publabel#1{\gdef\@publabel{#1}}
\long\def\author#1{\gdef\@author{#1}}
\long\def\title#1{\gdef\@title{#1}}
\long\def\abstract#1{\gdef\@abstract{#1}}
\def\titlerelax{
\let\maketitle\relax
\let\settitleparameters\relax
\let\consolidatetitle\relax
\let\inittitlepage\relax
\let\finishtitlepage\relax
\let\titlepagecontents\relax
\let\multithanks\relax
\let\titlebaselines\relax
\let\@makepub\relax
\let\@maketitle\relax
\let\@makeauthor\relax
\let\@makeabstract\relax
\let\@maketitlenote\relax
\let\thanks\relax
\let\titlerelax\relax}

\def\titleclean
{\gdef\@titlenote{}
\gdef\@abstract{}
\gdef\@author{}
\gdef\@title{}
\gdef\@pubdate{}\gdef\@pubnumber{}\gdef\@publabel{}
\gdef\@dpublabel{}
}
\def\@makepub{\vbox to \z@{\hbox to \textwidth{\hfill
\@publabel \hfill
\llap{\parbox[t]{0.25\textwidth}{\raggedleft\@pubnumber}}}%
\vss}}
\def\@maketitle{\vskip 60pt \begin{center}
 {\LARGE \@title \par}
 \end{center}}
\def\@makeauthor{{%
\def\and{\smallskip {\normalsize \rm and\smallskip }}
\def\And{\medskip {\normalsize \rm and\\}\medskip}
\long\def\address##1{{\def\and{\\and\\}\medskip
				{\small \it \\##1\\}
}}
{\centering
 \vskip 3em
 \large \lineskip .75em
 \@author}
 \par}} 
\def\@makedate{\vskip 1.5em 
 {\raggedright \small \noindent\@pubdate \par}}
\def\@makeabstract{\vskip 1.5em
{\small 
\begin{center}
{\bf ABSTRACT\vspace{-.5em}\vspace{0pt}} 
\end{center}
\quotation \@abstract \endquotation}}
\def\maketitle{\titlepage
\let\footnotesize\small \setcounter{page}{1}
\def\thefootnote{\arabic{footnote}}
\@makepub
\vfil
\@maketitle
\@makeauthor
\vfil
\@makeabstract
\@thanks
\vfil
\@makedate
\if@restonecol\twocolumn \else \eject \fi
\titlerelax \titleclean
\def\thefootnote{\alph{footnote}}
\setcounter{footnote}{0}
}
\ifcase\@ptsize
 \font\tenmsa=msam10
 \font\sevenmsa=msam7
 \font\fivemsa=msam5
 \font\tenmsb=msbm10
 \font\sevenmsb=msbm7
 \font\fivemsb=msbm5
\or
 \font\tenmsa=msam10 scaled \magstephalf
 \font\sevenmsa=msam8
 \font\fivemsa=msam6
 \font\tenmsb=msbm10 scaled \magstephalf
 \font\sevenmsb=msbm8
 \font\fivemsb=msbm6
\or
 \font\tenmsa=msam10 scaled \magstep1
 \font\sevenmsa=msam8
 \font\fivemsa=msam6
 \font\tenmsb=msbm10 scaled \magstep1
 \font\sevenmsb=msbm8
 \font\fivemsb=msbm6
\fi

\newfam\msafam
\newfam\msbfam
\textfont\msafam=\tenmsa  \scriptfont\msafam=\sevenmsa
  \scriptscriptfont\msafam=\fivemsa
\textfont\msbfam=\tenmsb  \scriptfont\msbfam=\sevenmsb
  \scriptscriptfont\msbfam=\fivemsb

\def\hexnumber@#1{\ifnum#1<10 \number#1\else
 \ifnum#1=10 A\else\ifnum#1=11 B\else\ifnum#1=12 C\else
 \ifnum#1=13 D\else\ifnum#1=14 E\else\ifnum#1=15 F\fi\fi\fi\fi\fi\fi\fi}

\def\msa@{\hexnumber@\msafam}
\def\msb@{\hexnumber@\msbfam}
\mathchardef\boxdot="2\msa@00
\mathchardef\boxplus="2\msa@01
\mathchardef\boxtimes="2\msa@02
\mathchardef\square="0\msa@03
\mathchardef\blacksquare="0\msa@04
\mathchardef\centerdot="2\msa@05
\mathchardef\lozenge="0\msa@06
\mathchardef\blacklozenge="0\msa@07
\mathchardef\circlearrowright="3\msa@08
\mathchardef\circlearrowleft="3\msa@09
\mathchardef\rightleftharpoons="3\msa@0A
\mathchardef\leftrightharpoons="3\msa@0B
\mathchardef\boxminus="2\msa@0C
\mathchardef\Vdash="3\msa@0D
\mathchardef\Vvdash="3\msa@0E
\mathchardef\vDash="3\msa@0F
\mathchardef\twoheadrightarrow="3\msa@10
\mathchardef\twoheadleftarrow="3\msa@11
\mathchardef\leftleftarrows="3\msa@12
\mathchardef\rightrightarrows="3\msa@13
\mathchardef\upuparrows="3\msa@14
\mathchardef\downdownarrows="3\msa@15
\mathchardef\upharpoonright="3\msa@16

\mathchardef\downharpoonright="3\msa@17
\mathchardef\upharpoonleft="3\msa@18
\mathchardef\downharpoonleft="3\msa@19
\mathchardef\rightarrowtail="3\msa@1A
\mathchardef\leftarrowtail="3\msa@1B
\mathchardef\leftrightarrows="3\msa@1C
\mathchardef\rightleftarrows="3\msa@1D
\mathchardef\Lsh="3\msa@1E
\mathchardef\Rsh="3\msa@1F
\mathchardef\rightsquigarrow="3\msa@20
\mathchardef\leftrightsquigarrow="3\msa@21
\mathchardef\looparrowleft="3\msa@22
\mathchardef\looparrowright="3\msa@23
\mathchardef\circeq="3\msa@24
\mathchardef\succsim="3\msa@25
\mathchardef\gtrsim="3\msa@26
\mathchardef\gtrapprox="3\msa@27
\mathchardef\multimap="3\msa@28
\mathchardef\therefore="3\msa@29
\mathchardef\because="3\msa@2A
\mathchardef\doteqdot="3\msa@2B

\mathchardef\triangleq="3\msa@2C
\mathchardef\precsim="3\msa@2D
\mathchardef\lesssim="3\msa@2E
\mathchardef\lessapprox="3\msa@2F
\mathchardef\eqslantless="3\msa@30
\mathchardef\eqslantgtr="3\msa@31
\mathchardef\curlyeqprec="3\msa@32
\mathchardef\curlyeqsucc="3\msa@33
\mathchardef\preccurlyeq="3\msa@34
\mathchardef\leqq="3\msa@35
\mathchardef\leqslant="3\msa@36
\mathchardef\lessgtr="3\msa@37
\mathchardef\backprime="0\msa@38
\mathchardef\risingdotseq="3\msa@3A
\mathchardef\fallingdotseq="3\msa@3B
\mathchardef\succcurlyeq="3\msa@3C
\mathchardef\geqq="3\msa@3D
\mathchardef\geqslant="3\msa@3E
\mathchardef\gtrless="3\msa@3F
\mathchardef\sqsubset="3\msa@40
\mathchardef\sqsupset="3\msa@41
\mathchardef\vartriangleright="3\msa@42
\mathchardef\vartriangleleft="3\msa@43
\mathchardef\trianglerighteq="3\msa@44
\mathchardef\trianglelefteq="3\msa@45
\mathchardef\bigstar="0\msa@46
\mathchardef\between="3\msa@47
\mathchardef\blacktriangledown="0\msa@48
\mathchardef\blacktriangleright="3\msa@49
\mathchardef\blacktriangleleft="3\msa@4A
\mathchardef\vartriangle="3\msa@4D
\mathchardef\blacktriangle="0\msa@4E
\mathchardef\triangledown="0\msa@4F
\mathchardef\eqcirc="3\msa@50
\mathchardef\lesseqgtr="3\msa@51
\mathchardef\gtreqless="3\msa@52
\mathchardef\lesseqqgtr="3\msa@53
\mathchardef\gtreqqless="3\msa@54
\mathchardef\Rrightarrow="3\msa@56
\mathchardef\Lleftarrow="3\msa@57
\mathchardef\veebar="2\msa@59
\mathchardef\barwedge="2\msa@5A
\mathchardef\doublebarwedge="2\msa@5B
\mathchardef\angle="0\msa@5C
\mathchardef\measuredangle="0\msa@5D
\mathchardef\sphericalangle="0\msa@5E
\mathchardef\varpropto="3\msa@5F
\mathchardef\smallsmile="3\msa@60
\mathchardef\smallfrown="3\msa@61
\mathchardef\Subset="3\msa@62
\mathchardef\Supset="3\msa@63
\mathchardef\Cup="2\msa@64

\mathchardef\Cap="2\msa@65

\mathchardef\curlywedge="2\msa@66
\mathchardef\curlyvee="2\msa@67
\mathchardef\leftthreetimes="2\msa@68
\mathchardef\rightthreetimes="2\msa@69
\mathchardef\subseteqq="3\msa@6A
\mathchardef\supseteqq="3\msa@6B
\mathchardef\bumpeq="3\msa@6C
\mathchardef\Bumpeq="3\msa@6D
\mathchardef\lll="3\msa@6E

\mathchardef\ggg="3\msa@6F

\mathchardef\circledS="0\msa@73
\mathchardef\pitchfork="3\msa@74
\mathchardef\dotplus="2\msa@75
\mathchardef\backsim="3\msa@76
\mathchardef\backsimeq="3\msa@77
\mathchardef\complement="0\msa@7B
\mathchardef\intercal="2\msa@7C
\mathchardef\circledcirc="2\msa@7D
\mathchardef\circledast="2\msa@7E
\mathchardef\circleddash="2\msa@7F
\def\ulcorner{\delimiter"4\msa@70\msa@70 }
\def\urcorner{\delimiter"5\msa@71\msa@71 }
\def\llcorner{\delimiter"4\msa@78\msa@78 }
\def\lrcorner{\delimiter"5\msa@79\msa@79 }
\def\yen{\mathhexbox\msa@55 }
\def\checkmark{\mathhexbox\msa@58 }
\def\circledR{\mathhexbox\msa@72 }
\def\maltese{\mathhexbox\msa@7A }
\mathchardef\lvertneqq="3\msb@00
\mathchardef\gvertneqq="3\msb@01
\mathchardef\nleq="3\msb@02
\mathchardef\ngeq="3\msb@03
\mathchardef\nless="3\msb@04
\mathchardef\ngtr="3\msb@05
\mathchardef\nprec="3\msb@06
\mathchardef\nsucc="3\msb@07
\mathchardef\lneqq="3\msb@08
\mathchardef\gneqq="3\msb@09
\mathchardef\nleqslant="3\msb@0A
\mathchardef\ngeqslant="3\msb@0B
\mathchardef\lneq="3\msb@0C
\mathchardef\gneq="3\msb@0D
\mathchardef\npreceq="3\msb@0E
\mathchardef\nsucceq="3\msb@0F
\mathchardef\precnsim="3\msb@10
\mathchardef\succnsim="3\msb@11
\mathchardef\lnsim="3\msb@12
\mathchardef\gnsim="3\msb@13
\mathchardef\nleqq="3\msb@14
\mathchardef\ngeqq="3\msb@15
\mathchardef\precneqq="3\msb@16
\mathchardef\succneqq="3\msb@17
\mathchardef\precnapprox="3\msb@18
\mathchardef\succnapprox="3\msb@19
\mathchardef\lnapprox="3\msb@1A
\mathchardef\gnapprox="3\msb@1B
\mathchardef\nsim="3\msb@1C
\mathchardef\napprox="3\msb@1D
\mathchardef\varsubsetneq="3\msb@20
\mathchardef\varsupsetneq="3\msb@21
\mathchardef\nsubseteqq="3\msb@22
\mathchardef\nsupseteqq="3\msb@23
\mathchardef\subsetneqq="3\msb@24
\mathchardef\supsetneqq="3\msb@25
\mathchardef\varsubsetneqq="3\msb@26
\mathchardef\varsupsetneqq="3\msb@27
\mathchardef\subsetneq="3\msb@28
\mathchardef\supsetneq="3\msb@29
\mathchardef\nsubseteq="3\msb@2A
\mathchardef\nsupseteq="3\msb@2B
\mathchardef\nparallel="3\msb@2C
\mathchardef\nmid="3\msb@2D
\mathchardef\nshortmid="3\msb@2E
\mathchardef\nshortparallel="3\msb@2F
\mathchardef\nvdash="3\msb@30
\mathchardef\nVdash="3\msb@31
\mathchardef\nvDash="3\msb@32
\mathchardef\nVDash="3\msb@33
\mathchardef\ntrianglerighteq="3\msb@34
\mathchardef\ntrianglelefteq="3\msb@35
\mathchardef\ntriangleleft="3\msb@36
\mathchardef\ntriangleright="3\msb@37
\mathchardef\nleftarrow="3\msb@38
\mathchardef\nrightarrow="3\msb@39
\mathchardef\nLeftarrow="3\msb@3A
\mathchardef\nRightarrow="3\msb@3B
\mathchardef\nLeftrightarrow="3\msb@3C
\mathchardef\nleftrightarrow="3\msb@3D
\mathchardef\divideontimes="2\msb@3E
\mathchardef\varnothing="0\msb@3F
\mathchardef\nexists="0\msb@40
\mathchardef\mho="0\msb@66
\mathchardef\thorn="0\msb@67
\mathchardef\beth="0\msb@69
\mathchardef\gimel="0\msb@6A
\mathchardef\daleth="0\msb@6B
\mathchardef\lessdot="3\msb@6C
\mathchardef\gtrdot="3\msb@6D
\mathchardef\ltimes="2\msb@6E
\mathchardef\rtimes="2\msb@6F
\mathchardef\shortmid="3\msb@70
\mathchardef\shortparallel="3\msb@71
\mathchardef\smallsetminus="2\msb@72
\mathchardef\thicksim="3\msb@73
\mathchardef\thickapprox="3\msb@74
\mathchardef\approxeq="3\msb@75
\mathchardef\succapprox="3\msb@76
\mathchardef\precapprox="3\msb@77
\mathchardef\curvearrowleft="3\msb@78
\mathchardef\curvearrowright="3\msb@79
\mathchardef\digamma="0\msb@7A
\mathchardef\varkappa="0\msb@7B
\mathchardef\hslash="0\msb@7D
\mathchardef\hbar="0\msb@7E
\mathchardef\backepsilon="3\msb@7F
\def\Bbb{\ifmmode\let\next\Bbb@\else
 \def\next{\errmessage{Use \string\Bbb\space only in math mode}}\fi\next}
\def\Bbb@#1{{\Bbb@@{#1}}}
\def\Bbb@@#1{\fam\msbfam#1}

\def\bk {{\hskip 0.2 cm}}

\def\com{{\hskip 0.2 cm},}
\def\pkt{{\hskip 0.2 cm}.}
\def\acknowledgements{\@startsection{section}{4}
{\z@}{-3.5ex plus -1ex minus -.2ex}{2.3ex plus .2ex}{\normalsize\bf}
{Acknowledgements}}

\def\half {\frac{1}{2}}           % 1/2
\def\thalf {\frac{3}{2}}          % 3/2
\newcommand{\tab}[1]{{\sc Tab.}\,{\sf #1}}        % Table reference
\newcommand{\eq}[1]{{\sc Eq.}\,{\sf (#1)}}        % Equation reference
\newcommand{\eqs}[1]{{\sc Eqs.}\,{\sf (#1)}}      % Equations reference
\newcommand{\refoth}[1]{{\sf #1}}                 % other references 
\newcommand{\eqoth}[1]{{\sf (#1)}}                % other equation references 

\def\bbbz {\Bbb{Z}}               % integers
\def\bbbzh{\Bbb{Z}_{\frac{1}{2}}} % Half integers
\def\bbbzu{\Bbb{Z}^{1/2}} % integers and Half integers

\def\bbbn {\Bbb{N}}               % +ve integers
\def\bbbnh{\Bbb{N}_{\frac{1}{2}}} % +ve half integers
\def\bbbno{\Bbb{N}_{0}}           % non -ve integers
\def\bbbnu{\Bbb{N}^{1/2}}
\def\bbbnuo{\Bbb{N}_0^{1/2}}
\def\bbbc {\Bbb{C}}

\newtheorem{definition}{Definition}[section]
\newtheorem{theorem}[definition]{Theorem}
\newtheorem{conjecture}[definition]{Conjecture}
\newtheorem{lemma}[definition]{Lemma}
\newtheorem{proposition}[definition]{Proposition}
\newcounter{defs}[section]

\newcommand{\be}{\begin{equation}}
\newcommand{\ee}{\end{equation}}
\newcommand{\bea}{\begin{eqnarray}}
\newcommand{\eea}{\end{eqnarray}}

\newcommand{\bdf}{\stepcounter{defs}\begin{definition}}
\newcommand{\edf}{\end{definition}}
\newcommand{\bth}{\stepcounter{defs}\begin{theorem}}
\newcommand{\eth}{\end{theorem}}
\newcommand{\bcj}{\stepcounter{defs}\begin{conjecture}}
\newcommand{\ecj}{\end{conjecture}}
\newcommand{\blm}{\stepcounter{defs}\begin{lemma}}
\newcommand{\elm}{\end{lemma}}
\newcommand{\bpr}{\stepcounter{defs}\begin{proposition}}
\newcommand{\epr}{\end{proposition}}
\newcommand{\bprf}{Proof: }
\newcommand{\eprf}{\hfill $\Box$ \\}
\newcounter{pics}
\newcommand{\bpic}[4]{\begin{center}\begin{picture}(#1,#2)(#3,#4)
\refstepcounter{pics}}
\renewcommand{\thepics}{{\sf\roman{pics}}}
\newcommand{\epic}[1]{\end{picture}\\
{\small {\sc Fig.} \thepics \bk #1} \end{center}}
\newcommand{\epicspl}{\end{picture}\\           % Different ending for picture
\addtocounter{pics}{-1}\end{center}}            % environment for splitt pictures
\renewcommand{\thefootnote}{\rm{\alph{footnote}}}
\newcounter{tabs}
\newcommand{\btab}[1]{\refstepcounter{tabs}\begin{center}
\begin{tabular}{#1}}
\renewcommand{\thetabs}{{\sf\alph{tabs}}}
\newcommand{\etab}[1]{\end{tabular}\\[1.5ex]
{\small {\sc Tab.} \thetabs \bk #1} \end{center}}

\def\noi {\noindent}
\newcommand{\nn}{\nonumber}
\newcommand{\ket}[1]{\left| {#1} \right\rangle} % ket
\newcommand{\spn}[1]{{\rm span}\{{#1}\}}        % span
\def\det{{\rm det}}                             % det
\newcommand{\vm}[1]{{\langle #1 \rangle}}       % Parametrised Verma modules
\def\pmb#1{\setbox0=\hbox{#1}%
 \kern-.025em\copy0\kern-\wd0
 \kern.05em\copy0\kern-\wd0
 \kern-.025em\raise.0433em\box0 }

\makeatother

\def\tw{{\sf J}_2}       % twisted N=2 algebra
\def\vm{{\cal V}}        % Verma module
\def\ordering{{\cal O}}  % ordering
\def\bsvm{{\cal B}}      % basis for Verma module                   
\def\cset{{\cal C}}      % order part C of basis
\def\sset{{\cal S}}      % suborder part S of basis 
\def\lset{{\sf L}}      % order part L of basis 
\def\tset{{\sf T}}      % order part T of basis 
\def\gset{{\sf G}}      % order part G^hat of basis 
\newcommand{\levelt}[1]{{{|}#1{|}_{L}}} % level of an operator in N=2
\newcommand{\charget}[1]{{{|}#1{|}_{F}}}        % parity of an operator in N=2
\newcommand{\length}[1]{{{\|}#1{\|}}}   % length of an operator
\newcommand{\osm}[1]{{{<}_{{}_{#1}}}}           % ordering smaller

\title{Singular dimensions of the $N=2$ superconformal algebras II: 
the twisted $N=2$ algebra}
\author{Matthias D\"{o}rrzapf\thanks{m.doerrzapf@damtp.cam.ac.uk}
and Beatriz Gato-Rivera\thanks{bgato@pinar1.csic.es}$^{,3}$
\address{$^1$Department of Applied Mathematics and Theoretical
Physics\\
University of Cambridge, Silver Street \\
Cambridge, CB3 9EW, UK\\[.3cm]
$^2$Instituto de Matem\'aticas y F\'\i sica Fundamental, CSIC,\\
Serrano 123, Madrid 28006, Spain \\[.3cm]
$^3$NIKHEF, Kruislaan 409, NL-1098 SJ Amsterdam, The Netherlands}}
\pubdate{February 1999}
\pubnumber{DAMTP-99-19\\ IMAFF-99-FM/08, NIKHEF-99-06\\
hep-th/9902044 }

\abstract{We introduce a suitable {\it adapted ordering} for the twisted 
$N=2$ superconformal algebra (i.e. with mixed boundary conditions for the 
fermionic fields). We show that the ordering kernels for complete Verma 
modules have two elements and the ordering kernels for $G$-closed Verma 
modules just one. Therefore, spaces of singular vectors may be 
two-dimensional for complete Verma modules whilst for $G$-closed Verma 
modules they can only be one-dimensional. We give all singular vectors for 
the levels $\half$, $1$, and $\thalf$ for both complete Verma modules and 
$G$-closed Verma modules. We also give explicit examples of degenerate cases 
with two-dimensional singular vector spaces in complete Verma modules. 
General expressions are conjectured for the relevant terms of all (primitive) 
singular vectors, i.e. for the coefficients with respect to the ordering 
kernel. These expressions allow to identify all degenerate cases as well as 
all $G$-closed singular vectors. They also lead to the discovery of 
subsingular vectors for the twisted $N=2$ superconformal algebra. Explicit 
examples of these subsingular vectors are given for the levels $\half$, $1$, 
and $\thalf$. Finally, the multiplication rules for singular vector operators 
are derived using the ordering kernel coefficients. This sets the basis for 
the analysis of the twisted $N=2$ embedding diagrams.}

\begin{document}

\maketitle

%\tableofcontents

%%%%%%%%%%%%%%%%%%%%%%%%%%%%%%%%%%%%%%%%%%%%%%%%%%%%%%%%%%%
%                                                         %
%                 Introduction                            %
%                                                         %
%%%%%%%%%%%%%%%%%%%%%%%%%%%%%%%%%%%%%%%%%%%%%%%%%%%%%%%%%%%

\section{Introduction}

The spectacular gain of importance of superstring theory in physics over 
the past fifteen years
asked for solutions to challenging problems on the mathematics side. 
Just like for any quantum field theory, a 
key problem is the study of the underlying symmetry algebra in order 
to understand the space of states
of the physical theory. Almost at the same time as
Ademollo et al.\cite{ademollo} identified for the first time
the symmetry algebras of the superstrings, Kac was
already studying these objects among his classification
of superalgebras\cite{kac3}.  These symmetry algebras are known as 
superconformal algebras because the Virasoro algebra 
is a subalgebra due to the conformal invariance on the string
world-sheet. The study of the 
representations of the superconformal algebras is one of these
challenging problems for mathematics. 
Despite its importance, 
satisfactory answers are still not known for most superconformal
algebras. Increasing the number of fermionic currents $N$
the structure of the representations of the superconformal algebras gets
more and more complicated. The case of the $N=1$ superconformal 
algebra is well understood thanks to many 
authors\cite{ast1,bsa2,bsa3,friedan,fuchs,gko,kac4,kiritsis2,meur,gerard}. 
However, its representation theory is
very close to the representation theory of the Virasoro algebra. Already 
for the $N=2$ superconformal algebras there are still many open 
questions, even though much progress has been made in recent 
years\cite{bfk,cmp1,thesis,beatriz1,beatriz2,npb1}. 
For the $N=2$ superconformal algebras many aspects arise that are new
to representations of chiral algebras.
One of these aspects is the existence of degenerate singular 
vectors\cite{cmp1,beatriz2}. That is,
$N=2$ superconformal Verma modules can have at the same 
level and with the same charge more than one linearly 
independent singular vectors.
Thus the study of the embedding structure as well as the computation 
of character formulae for the 
representations is much more complicated than in the case where the 
comparison of level and charge
is already enough in order to decide if two singular vectors are
proportional, as it is for the Virasoro
algebra and for the $N=1$ superconformal algebra.

In physics, the different types of superconformal symmetry algebras, 
for the same number of fermionic fields $N$, 
arise through the different choices of periodicity conditions for the 
fermionic currents around a closed superstring world-sheet, as well as
for different choices of the Virasoro generators (i.e. of the stress-energy
tensor). For the standard
choice of Virasoro generators, the superconformal algebras resulting from
antiperiodic boundary conditions are called Neveu-Schwarz algebras, 
those corresponding to periodic boundary conditions are
called Ramond algebras and those with mixed boundary conditions are called 
twisted superconformal algebras. 
The $N=2$ superconformal algebras fall under four
types: the Neveu-Schwarz $N=2$ algebra, the Ramond $N=2$ algebra,
the topological $N=2$ algebra, and the twisted $N=2$ algebra. The
first three algebras are isomorphic, nevertheless their highest 
weight representations are quite different\footnote{We thank V. Kac for 
discussions on this point}. 
The topological $N=2$ algebra, which is the
symmetry algebra of topological conformal fied theory, also plays an 
important r\^ole in string theories\cite{dvv,grs,blnw}. This 
algebra is also known as the `twisted topological' $N=2$ algebra because 
it can be obtained from the Neveu-Schwarz $N=2$ algebra by modifying the
stress-energy tensor by adding the derivative of the U(1) current, 
procedure known as `topological twist'. 

We have studied the singular dimensions of 
the three isomorphic algebras in a recent publication\cite{p6sdim1}, 
where we presented a formalism - the adapted ordering method -
that allows us to compute upper limits
for the dimensions of singular subspaces simply by ordering the
algebra generators appropriately. If this order is chosen in a suitable 
way, then the computed upper limits for
these singular dimensions may actually be maxima. We applied this 
method directly to the topological $N=2$ algebra proving its usefullness, 
and then we deduced the maximal singular dimensions corresponding to 
the Neveu-Schwarz and to the Ramond $N=2$ algebras. 

So far, not much attention has been given to the twisted superconformal 
algebras, which appear for $N=2$ for the first time. This is mainly due to 
the fact that field theories with mixed periodicity conditions have not 
been considered for string theories. 
In the $N=2$ case there is just one twisted algebra
whilst for bigger $N$ there are several possible ways of mixing the 
periodicity conditions of
the fermionic fields and thus different twisted superconformal
algebras can be found. 
In this paper we will focus on the twisted $N=2$ superconformal algebra. 
We will show that the method of adapted orderings can also be used for this
algebra. For the three isomorphic $N=2$ algebras as well 
as for the Virasoro algebra
we saw\cite{p6sdim1} that the suitably chosen adapted ordering assigns 
a special r\^ole to the powers of the
Virasoro generator $L_{-1}$. Surprisingly enough, in the 
twisted $N=2$ case this r\^ole is taken over for the first time
by another generator. Furthermore, we will see that the twisted $N=2$ 
algebra requires its two fermionic fields to be mixed rather
than kept separate as for the other $N=2$ algebras. Hence the 
twisted $N=2$ algebra behaves very differently from the other $N=2$ 
algebras with respect to the adapted ordering method. Nevertheless, 
the method still works and we can prove that the computed upper limits 
for the singular dimensions are actually maxima. 
Even though the twisted $N=2$ superconformal algebra contains two 
fermionic fields, at the algebraic level it looks as if it really 
contains only one. Furthermore the parameter space 
of the conformal weights is just one-dimensional, unlike the corresponding
parameter spaces for the three isomorphic $N=2$ algebras. 
For both reasons, one would have naively
suggested that the singular dimensions have to be less than or equal 
to $1$. Surprisingly enough
we will show that the maximal singular dimension is in fact $2$. Surely, 
these two-dimensional spaces are not spanned by any tangent 
space of vanishing surfaces corresponding to singular
vectors - as it is the case
for the three isomorphic $N=2$ algebras - simply due to the fact that 
the parameter space is
only one-dimensional. The singular dimension $2$ arises rather in a 
completely new way due to the 
intersection of two different one-parameter families of singular vectors.  
This is another reason for which the twisted $N=2$ algebra is so far
unique of its kind.

Superconformal algebras seem to have many surprising features. Not only 
does this paper show that recent achievements follow 
the correct way to tackle these and to learn
much, which will be helpful to analyse even more complicated conformal 
structures, it also shows that 
the twisted $N=2$ algebra gives more new and different insight in 
superconformal representation theory. Furthermore
this paper creates the basis for the study of the embedding structure 
of the twisted $N=2$ highest weight representations since 
we deduce the singular dimensions as well as the
multiplication rules for singular vector operators. 
Both are crucial for the analysis of the embedding structure.
We will also give answers to the question whether the twisted $N=2$
superconformal algebra contains subsingular vectors. The study of
the embedding diagrams, that is the structure of descendant (secondary) 
singular vectors, will be considered in a forthcoming publication.

This paper is organised in the following way. After a small introduction 
with the necessary facts about
the twisted $N=2$ algebra in section \refoth{\ref{sec:alg}}, we review  
the concept of adapted orderings and its 
main implications in section \refoth{\ref{sec:ordering}}, 
in a version suitable for the twisted $N=2$ algebra. 
In section \refoth{\ref{sec:adapted}}
we then introduce an adapted ordering for the Verma modules of the 
twisted $N=2$ algebra.
This allows us to compute the upper limits for the singular dimensions 
and the multiplication rules for singular vector 
operators in section \refoth{\ref{sec:sdim}}. That these  
upper limits are in fact maximal dimensions 
is demonstrated in section \refoth{\ref{sec:examples}} 
using explicit examples which are supplemented by more
examples in appendix \refoth{\ref{app:a}}. 
Based on both theoretical results and explicit computational
results we give a reliable conjecture for the coefficients of the
relevant terms of all twisted $N=2$ singular vectors in section 
\refoth{\ref{sec:classification}}. 
Using this conjecture we can characterize all cases of degenerate 
(i.e. two-dimensional) singular vectors for all levels. This conjecture 
leads to the discovery of subsingular vectors, which are singular in the
$G$-closed Verma modules considered in section \refoth{\ref{sec:GclosedVM}}
and in appendix \refoth{\ref{app:b}}. It also leads to the identification
of all $G$-closed singular vectors, which are
the issue of section \refoth{\ref{sec:Gclosedsvecs}}. We conclude the 
paper with some final remarks
in section \refoth{\ref{sec:conclusions}}.

%%%%%%%%%%%%%%%%%%%%%%%%%%%%%%%%%%%%%%%%%%%%%%%%%%%%%%%%%%%
%                                                         %
%  the twisted N=2 algebra (mixed periodicity conditions)    %
%                                                         %
%%%%%%%%%%%%%%%%%%%%%%%%%%%%%%%%%%%%%%%%%%%%%%%%%%%%%%%%%%%

\section{The twisted $N=2$ superconformal algebra}
\label{sec:alg}

The twisted $N=2$ superconformal algebra $\tw$ consists of the Virasoro
algebra generators\footnote{We use the notation $\bbbn=\{1,2,3,\ldots\}$, 
$\bbbno=\{0,1,2,\ldots\}$,
$\bbbnh=\{\frac{1}{2},\frac{3}{2},\frac{5}{2},\ldots\}$, 
$\bbbnu=\bbbn\cup\bbbnh$, $\bbbnuo=\bbbno\cup\bbbnh$, and also 
$\bbbz=\{\ldots,-1,0,1,2,\ldots\}$, 
$\bbbzh=\{\ldots,-\frac{3}{2},-\frac{1}{2},\frac{1}{2},\frac{3}{2},\ldots\}$,
$\bbbzu=\bbbz\cup\bbbzh$.} $L_{m}$, 
$m\in\bbbz$, corresponding to the stress-energy tensor, a Heisenberg 
algebra $T_r$, with half-integral $r\in\bbbzh$, corresponding to the
U(1) current, and the fermionic generators $G_k$, 
$k\in\bbbzu$, which are the modes of the two spin-3/2 fermionic fields. 
$\tw$ satisfies the following commutation
relations\footnote{The earlier way of writing the twisted $N=2$ 
commutation relations, as for instance
given by Boucher, Friedan, and Kent\cite{bfk}, is to
distinguish the modes of the two fermionic fields explicitly using 
superscripts $1$ and $2$.
However, algebraically it is not necessary to distinguish them 
as one of the fields has modes with integral indices and the other one 
has modes with half-integral indices.
In order to avoid the imaginary unit $i$ appearing in the 
commutation relations,
we have, for convenience, redefined $iG_r$ as $G_r$ for $r\in\bbbzh$ 
compared to the notation in Ref. \icite{bfk}.}.
\bea
[L_{m},L_{n}] & = & (m-n) L_{m+n} + \frac{C}{12} \:(m^{3}-m)\:
\delta_{m+n,0} \;\; ,\nn \\
\ [L_{m},G_{k}] & = & (\frac{1}{2} m-k) G_{m+k} \;\; ,\nn \\
\ [L_{m},T_{r}] & = & -r T_{m+r} \;\; ,\nn \\
\ [T_{r},T_{s}] & = & \frac{1}{3} C r \delta_{r+s,0}  
\;\; ,\label{eq:cr} \\
\ [T_{r},G_{k}] & = & G_{r+k} \;\; ,\nn \\
\ \{ G_{k},G_{l}\} & = & 2 \delta^{\bbbz}_{k+l} (-1)^{2k} L_{k+l}- 
\delta^{\bbbzh}_{k+l} (-1)^{2k} (k-l) T_{k+l}
 +\frac{C}{3} \delta^{\bbbz}_{k+l} (-1)^{2k} (k^{2}-\frac{1}{4}) 
 \delta_{k+l,0} \;\; ,\nn \\
\ [L_{m},C] & = & [T_{r},C] \;\; = \;\; [G_{k},C]  \;\; = \;\; 0 \;\; , \nn
\eea
with $\delta^{S}_m=1$ if $m\in S$ and 
$\delta^{S}_m=0$ otherwise, $(-1)^{2k}=+1$ for $k\in\bbbz$ and
$(-1)^{2k}=-1$ for $k\in\bbbzh$, $m,n\in\bbbz$, $r,s\in\bbbzh$, and
finally $k,l\in\bbbzu$. For the Neveu-Schwarz, Ramond and topological 
$N=2$ algebras one usually chooses a basis in which the
odd generators are nilpotent. If we want
to keep the $L_0$-grading of the basis for the algebra generators
then such a choice does not exist for the twisted $N=2$ algebra.
Nevertheless, the squares of the fermionic operators
can be expressed in terms of Virasoro operators, as one can deduce 
easily from the commutation relations \eqs{\ref{eq:cr}} :
\bea
G_k^2 &=& (-1)^{2k} L_{2k} - \delta_{k,0}\frac{C}{24} \com k\in\bbbzu 
\pkt \label{eq:square} 
\eea

The central term $C$ commutes with all other operators and can therefore
be fixed as $c\in\bbbc$. ${\cal H}_{\tw}=\spn{L_0, C}$ 
defines a commuting subalgebra
of $\tw$, which can therefore be 
diagonalised simultaneously. Generators with positive
index span the set of {\it positive operators} $\tw^+$ of $\tw$
and likewise generators with negative index 
span the set of {\it negative operators} $\tw^-$ of $\tw$:
\bea
\tw^{+} &=& \spn{L_{m},T_{r},G_{k}: m\in\bbbn,r\in\bbbnh,k\in\bbbnu} \,, \\
\tw^{-} &=& \spn{L_{-m},T_{-r},G_{-k}: m\in\bbbn,r\in\bbbnh,k\in\bbbnu} \,.
\eea 
\noi The {\it zero modes} are spanned by $\tw^0=\spn{L_0,G_0,C}$ 
such that the generator $G_0$ classifies the 
different types of Verma modules as we will analyse shortly. 
One usually extends the algebra by the parity operator\footnote{Note 
that due to \eq{\ref{eq:square}}
the fermion number $F$ is not well defined, however, the {\it 
parity} $(-1)^F$ is.} $(-1)^F$ which commutes
with all operators $L_m$, $T_r$, and $C$ and anticommutes with $G_k$:
\bea
[(-1)^F,L_m] = [(-1)^F,T_r]=[(-1)^F,C] &=& 0 \com m\in\bbbz \,, 
r\in\bbbzh \,, \nn \\
\{(-1)^F,G_k\} &=& 0 \com k\in\bbbzu \,. \nn
\eea
The operator $(-1)^F$ will serve us later to distinguish fermionic from 
bosonic states in the space of states.
As $(-1)^F$ commutes with ${\cal H}_{\tw}$ it can also be diagonalised 
simultaneously to ${\cal H}_{\tw}$.

As usual, a simultaneous eigenvector $\ket{\Delta}$ of
${\cal H}_{\tw}$ and $(-1)^F$ with $L_0$-eigenvalue $\Delta$
(the conformal weight), $C$-eigenvalue\footnote{For
simplicity we will supress the eigenvalue of $C$ in 
$\ket{\Delta,c}$ and simply write $\ket{\Delta}$.}
$c$ (the conformal anomaly), and vanishing 
$\tw^{+}$ action is called a {\it highest weight vector}, corresponding
in fact to the state with {\it lowest} conformal weight in a given
representation of the algebra. Unless otherwise stated, we set
$\ket{\Delta}$ to have $(-1)^F$-eigenvalue\footnote{Instead of parity $\pm 1$ we shall
simply say parity $\pm$.} (parity) $+$. Additional zero-mode 
vanishing conditions are possible only with respect to the operator $G_0$ 
which may or may not annihilate a highest weight vector $\ket{\Delta}$
(c.f. Ref. \icite{p6sdim1}, definition \refoth{3.A}). If $G_0$ annihilates
the highest weight vector we shall denote it by a superscript
in $\ket{\Delta}^{G}$ and call this highest weight vector $G${\it-closed}. 
Since $G_0^2=L_0-\frac{c}{24}$, any $G$-closed vector necessarily has 
conformal weight $\Delta=\frac{c}{24}$. Verma modules are 
defined in the usual way
as the left module obtained by acting with the
universal enveloping algebra $U(\tw)$ on a highest weight vector: 
\bea
\label{eq:vms}
\vm_{\Delta} &=& U(\tw)\otimes_{{\cal H}_{\tw}\oplus\tw^+}\ket{\Delta} \,,\\
\vm_{\frac{c}{24}}^G &=& U(\tw)\otimes_{{\cal H}_{\tw}\oplus\tw^+\oplus
\spn{G_0}}\ket{\frac{c}{24}}^G \,,
\eea
If the highest weight vector is $G$-closed, then we call the Verma module
$\vm_{\frac{c}{24}}^G$ (which is not complete therefore) also $G$-closed. 

For an eigenvector $\Psi^p_{l}$ of ${\cal H}_{\tw}$  
in $\vm_{\Delta}$ the conformal weight is $\Delta+l$
and the parity is $p$ with $l\in\bbbnuo$ and $p\in\{\pm\}$. For convenience
we will sometimes denote the {\it level} and the {\it parity} as
$\levelt{\Psi^p_{l}}=l$ and $\charget{\Psi^p_{l}}=p$. {\it Singular vectors} 
$\Psi^{\pm}_{l}$ of a Verma module $\vm_{\Delta}$ are
eigenvectors of ${\cal H}_{\tw}$ with parity $\pm$ that are not 
proportional to the highest weight vector but are annihilated by
$\tw^+$. They correspond therefore to the states with lowest conformal 
weight in a given subrepresentation of the algebra.  
They may also satisfy an additional vanishing condition with 
respect to the operator $G_0$ in which case we add a superscript $G$ in 
the notation
$\Psi^{\pm G}_{l}$ and call it a {\it $G$-closed singular vector}. 
Obviously $G$-closed singular vectors satisfy similar restrictions on their
conformal weight as $G$-closed highest weight vectors:
$\Delta+l=\frac{c}{24}$. Most singular vectors can be constructed by acting
with algebra generators on another singular vector at a lower (or equal)
level. These are called {\it secondary} singular vectors. Otherwise, they 
are called {\it primitive} singular vectors.
The {\it singular vector operator}
$\theta_{l}^{\pm}$ of $\Psi^{\pm}_l$ is the unique operator in
$U(\tw^-\oplus\{G_0\})$ such that
$\Psi^{\pm}_l=\theta^{\pm}_l\ket{\Delta}$ and
similarly for $\Psi^{\pm G}_l$ where $U(\tw^-\oplus\{G_0\})$ is replaced
by $U(\tw^-)$.

Singular vectors and all their descendant vectors (i.e. all the states 
that are produced by acting with
algebra generators on the singular vectors) have vanishing 
{\it (pseudo-)}norms under the standard
inner product. Vectors with zero norm are usually called 
{\it null vectors}\footnote{Strictly speaking,
{\it null-vectors} should only be vectors in the kernel of the 
inner product matrix and hence decouple
completely from the whole space of states. If the pseudo-norm is 
not positive semi-definite 
there may be vectors with vanishing pseudo-norm that are not 
null-vectors\cite{p11vanish}.}.
The quotient module where
all null vectors are set to $0$ is hence an irreducible highest weight 
representation.
Conversely, all irreducible highest weight representations can be 
constructed in this way.
However, singular vectors and their descendants do not necessarily span the
whole submodule of null vectors. The quotient module of the Verma 
module divided by
the submodule spanned by all singular vectors may again contain new singular
vectors, called subsingular vectors. Subsingular vectors and their 
descendants are
also null vectors. Subsingular vectors do not appear for the Virasoro 
algebra neither for
the $N=1$ superconformal algebras. However, they have been discovered 
for the three isomorphic $N=2$
superconformal algebras in Refs. \icite{beatriz2,beatriz1}. 
In these cases, the existence of subsingular vectors is very closely
related to the existence of singular operators annihilating a singular
vector
as shall be discussed in a forthcoming publication\cite{p11vanish}. 
That the twisted $N=2$ superconformal algebra
also has subsingular vectors will be shown in section
\refoth{\ref{sec:GclosedVM}}. of this paper.

Even though the set of null vectors consists of singular vectors, 
subsingular vectors, and all their descendants,
the null vectors with lowest level in a Verma module are certainly  
singular vectors (i.e. annihilated by all the positive operators). 
Therefore, the lowest level at which the determinant of the inner product 
matrix vanishes indicates  the presence of (at least) one singular
vector. Hence the determinant formula is one of the most important
tools for analysing highest weight representations.
For the twisted $N=2$ superconformal algebra the determinant
formula has been given by Boucher, Friedan, and Kent\cite{bfk}. For
level zero the determinant formulae are $\det{M}^+_0=1$ and 
$\det{M}^-_0= \Delta-\frac{c}{24}$. For level $l\in\bbbnu$ and parity $\pm$ 
the determinant formulae are:
\bea
\det{M}^{\pm}_l &=& (\Delta-\frac{c}{24})^{\frac{P(l)}{2}} 
\prod_{1\leq rs\leq 2l \atop s {\rm odd}}
[2(\frac{c}{3}-1)(\Delta-\frac{c}{24})+
\frac{1}{4}((\frac{c}{3}-1)r+s)^2]^{P(l-\frac{rs}{2})} \com \label{detform}
\eea
where $r$ and $s$ are positive integers ($s$ odd). The partition function 
$P(l)$, that can be found in Ref. \icite{bfk}, satisfies $P(0)=1$.
The two parity sectors $\pm$ have exactly the same determinant expressions. 
This correspondence
is in fact even closer since the operator $G_0$ interpolates between the 
two sectors, provided
$G_0$ does not act on any $G$-closed vector. Let us analyse this 
correspondence closer for singular vectors.

For a given singular vector 
$\Psi^+_{l}=\theta^+_l\ket{\Delta}\in\vm_{\Delta}$ with singular operator 
$\theta^+_l$ we can construct two negative parity singular vectors:
\bea
\Psi_{G,l}^- &=& \theta^+_lG_0\ket{\Delta} \com \label{eq:psiG1} \\ 
{}_G\Psi_{l}^- &=& G_0\theta^+_l\ket{\Delta} \pkt
\eea
Likewise, $G_0$ constructs two positive parity singular vectors from a 
singular vector
$\Psi^-_{l}=\theta^-_l\ket{\Delta}\in\vm_{\Delta}$:
\bea
\Psi_{G,l}^+ &=& \theta^-_lG_0\ket{\Delta} \com \\
{}_G\Psi_{l}^+ &=& G_0\theta^-_l\ket{\Delta} \pkt \label{eq:psiG4}
\eea
One might therefore believe that singular vectors always come in groups 
of (at least) $4$. However, this assumes that these $4$ vectors 
are actually linearly independent, what is not true as one can guess  
from the fact that $P(0)=1$ in the determinant formulae, pointing towards
the existence of, generically, one singular vector of each parity at the
same level in the same Verma module. This issue will be considered in 
section \refoth{\ref{sec:classification}}. In the case of $G$-closed 
Verma modules or $G$-closed singular
vectors some of these vectors are actually trivial, an issue that is 
important for the discussion of embedding diagrams.

For elements $Y$ of $\tw$ which are eigenvectors of ${\cal H}_{\tw}$ 
and $(-1)^F$ with respect
to the adjoint representation we define
the level $\levelt{Y}$ as $[L_0,Y]=\levelt{Y}Y$ and in
addition the parity $\charget{Y}$ as $[(-1)^F,Y]=\charget{Y}Y$.
In particular, elements of the form
\bea 
Y&=&L_{-m_L} \ldots L_{-m_1}T_{-s_T} \ldots T_{-s_1} 
G_{-k_G} \ldots G_{-k_1}T_{-\frac{1}{2}}^n G_{-\frac{1}{2}}^{r_1}
G_{0}^{r_2} ,
\eea
\noi where $r_1,r_2\in\{0,1\}$ and any reorderings of $Y$ have 
level $\levelt{Y}=\sum_{j=1}^{L}m_j
+\sum_{j=1}^{T}s_j+\sum_{j=1}^{G}k_j+\frac{n+r_1}{2}$ and
parity $\charget{Y}=(-1)^{G+r_1+r_2}$. For these elements we shall also
define their {\it length} $\length{Y}=L+T+G$. For the trivial case we set
$\levelt{1}=\length{1}=0$ and $\charget{1}=+1$.
For convenience we define the following sets of negative
operators for $l\in\bbbnu$
\bea
\lset_l &=& \{ Y=L_{-m_L} \ldots L_{-m_1}: \; m_L\geq\ldots\geq m_1\geq 1,
\; \levelt{Y}=l\} \,, \\
\tset_l &=& \{ Y=T_{-s_T} \ldots T_{-s_1}: \; s_T\geq\ldots \geq s_1\geq 
\thalf , \; \levelt{Y}=l\} \,, \\
\gset_l &=& \{ Y=G_{-k_G} \ldots G_{-k_1}: \; k_G > \ldots > k_1\geq 1,
\; \levelt{Y}=l\} \,, \\
\lset_0 &=& \tset_0 ~=~ \gset_0 ~=~ \{1\} \,.
\eea
We can now define a graded basis for the Verma modules
which we use as the standard basis and on which we shall later 
construct the adapted ordering. 
A preliminary step is to define for $l\in\bbbnuo$
\bea
\sset^{\pm}_{l} &=& 
\left\{Y=LTG: \; L\in\lset_m, \; T\in\tset_r, \; G\in\gset_k, 
\right. \nn
\\ 
&& \left. \levelt{Y}=l=m+r+k, \;\charget{Y}=\pm 1=\charget{G} ,\; 
\,m,\in\bbbno ,\, r,k\in\bbbnuo\right\} \,.
\eea
This leads for $l\in\bbbnuo$ to:
\bea
\cset^{\pm}_{l} 
&=& \left\{S_{k,p} \, T_{-\frac{1}{2}}^{2l-2k-r_1}
G_{-\frac{1}{2}}^{r_1}G_{0}^{r_2}: \;S_{k,p}\in\sset^p_{k},
\;k\in\bbbnuo,\;r_1,r_2\in\{0,1\}, \right. \nn \\
&& \left. 2l-2k-r_1\geq 0, \, p(-1)^{r_1+r_2}=\pm 1 \right\} \,.  
\label{eq:cset} 
\eea
A typical element of $\cset^{\pm}_{l}$ is hence of the form 
\bea
Y &=& L_{-m_L}\ldots L_{-m_1}T_{-s_T}\ldots T_{-s_1}
G_{-k_G}\ldots G_{-k_1}
T_{-\frac{1}{2}}^{n} G_{-\frac{1}{2}}^{r_1} G_{0}^{r_2} \,,
\eea
$r_1,r_2\in\{0,1\}$, such that
$\levelt{Y}=m$ and $\charget{Y}=\pm 1$.
$S_{k,p}$ of $Y\in\cset^{\pm}_{l}$ is
called the {\it leading part of} $Y$ and is denoted by $Y^{\ast}$.
Hence, we can define the following standard basis:
\bea
{\cal B}_{\Delta} &=& \left\{ X\ket{\Delta}: 
\; X\in\cset^{\pm}_{l} \;, l\in\bbbnuo \; \right\} \,,
\label{eq:basis1}
\eea
\noi 
obtaining finally the Verma module $\vm_{\Delta}=\spn{{\cal B}_{\Delta}}$.

The basis \eq{\ref{eq:basis1}} is naturally 
$\bbbnuo\otimes\{\pm 1\}$ graded with respect to their $L_0$ and $(-1)^F$
eigenvalues, where the $L_0$-eigenvalue (the level) is seen relative to 
the eigenvalue $\Delta$ of the highest weight vector. 

The {\it normal form} of an eigenvector of the commuting subalgebra
${\cal H}_{\tw}$ is defined as the basis decomposition
with respect to the standard basis \eq{\ref{eq:basis1}}.
\bea
\Psi^p_{l} &=& \sum_{X\in\cset^p_{l}} c_X X\ket{\Delta} \pkt
\eea 
We call the operators 
$X\in\cset^p_{l}$ simply the {\it terms} of 
$\Psi^p_{l}$ and the coefficients $c_X$ its {\it coefficients}. 
Terms $X$ with non-trivial coefficients
$c_X$ are called {\it non-trivial terms} of $\Psi^p_{l}$.

For each of the two types of twisted Verma modules $\vm_{\Delta}$ 
and $\vm^G_{\frac{c}{24}}$ we
can thus think of two types of singular vectors $\Psi^{\pm}_{l}$ 
and $\Psi^{\pm G}_{\frac{c}{24}-\Delta}$ (both).
However, the latter does not exist in $G$-closed Verma modules, as
$G$-closed Verma modules $\vm_{\frac{c}{24}}$ contain at level 
$0$ only the highest weight vector\footnote{This resembles very much 
the result, for the Neveu-Schwarz and the topological N=2 algebras, 
that chiral singular vectors do not exist in chiral Verma modules.} .

To conclude this section, we should mention that in the case of a 
$G$-closed Verma module  
$\vm^G_{\frac{c}{24}}$ the basis \eq{\ref{eq:basis1}} simplifies to 
\bea
{\cal B}^G_{\frac{c}{24}} &=& \left\{ X\ket{\frac{c}{24}}: 
\; X\in\cset^{\pm G}_{l} \;, l\in\bbbnuo, \; \right\} \,,
\label{eq:basisG}
\eea
\noi where
\bea
\cset^{\pm G}_{l} 
&=& \left\{S_{k,p} \, T_{-\frac{1}{2}}^{2l-2k-r_1}
G_{-\frac{1}{2}}^{r_1}: \;S_{k,p}\in\sset^p_{k},
\;k\in\bbbnuo,\;r_1\in\{0,1\}, \right. \nn \\
&& \left. 2l-2k-r_1\geq 0, \, p(-1)^{r_1}=\pm 1 \right\} \,, l\in\bbbnuo \,.
\label{eq:csetG} 
\eea

\noi
All other definitions and results carry over to the $G$-closed case simply 
by considering $\Delta=\frac{c}{24}$.

%%%%%%%%%%%%%%%%%%%%%%%%%%%%%%%%%%%%%%%%%%%%%%%%%%%%%%%%%%%
%                                                         %
%          adapted orderings and singular dimensions      %
%                                                         %
%%%%%%%%%%%%%%%%%%%%%%%%%%%%%%%%%%%%%%%%%%%%%%%%%%%%%%%%%%%

\section{Adapted orderings and singular dimensions}
\label{sec:ordering}

A key question for the theory of highest weight representations 
is the dimension of any space of singular vectors that may appear 
at a given level $l\in\bbbnuo$ with fixed 
parity $p\in\{\pm\}$ in a Verma module $\vm_{\Delta}$.
In Ref. \icite{p6sdim1} we have shown that these dimensions, called singular
dimensions, are closely related to the ordering kernel of an adapted ordering
on the set $\cset^p_{l}$. In fact, it is the number of elements of the 
ordering kernel that sets an upper limit on the singular 
dimensions of a Verma module. Before
we introduce an adapted ordering $\ordering$ on $\cset^{\pm}_{l}$ and
compute its ordering kernel as defined in Ref. \icite{p6sdim1}, we shall 
first review
a few results on adapted orderings that are crucial for our reasoning. 
An adapted ordering is simply
a one-sided bounded total ordering on $\cset^p_{l}$ 
satisfying the requirements specified in the following definition.  

\bdf \label{def:adapt}
A total ordering $\ordering$ on $\cset^p_{l}$ with global  
minimum is called adapted to the
subset $\cset^{pA}_{l}\subset\cset^p_{l}$, 
in the Verma module $\vm_{\Delta}$
with annihilation operators ${\cal K}=\tw^+$,
if for any element $X_0\in\cset^{pA}_{l}$ at least one 
annihilation operator
$\Gamma \in{\cal K}$ exists for which
\bea
\Gamma \, X_0\ket{\Delta} &=&
\sum_{X\in \bsvm_{\Delta}} c_X^{\Gamma X_0} \, X  
\label{eq:adapt1}
\eea
contains a non-trivial term $\tilde{X}\in\bsvm_{\Delta}$ 
(i.e. $c_{\tilde{X}}^{\Gamma X_0}\neq0$) such
that for all $Y\in\cset^p_{l}$ 
with $X_0 \osm{\ordering} Y$ and $X_0\neq Y$ the coefficient
$c_{\tilde{X}}^{\Gamma Y}$ in
\bea
\Gamma \, Y\ket{\Delta} &=&
\sum_{X\in \bsvm_{\Delta}} c_X^{\Gamma Y} \, X  
\eea
 is trivial: $c_{\tilde{X}}^{\Gamma Y}=0$.
The complement of $\cset^{pA}_{l}$, 
${\ }\cset^{pK}_{l}=\cset^p_{l}\setminus
\cset^{pA}_{l}$ is the kernel with respect to
the ordering $\ordering$ in the Verma module $\vm_{\Delta}$.
%Here $\bsvm_{\Delta}$ represents a basis
%that can be chosen suitably for each $X_0$ and may or may not be the
%standard basis\footnote{In the $N=2$ case we will
%choose for most $X_0$ the standard basis
%with only very few but important exceptions.}. 
\edf

If we replace in definition \refoth{\ref{def:adapt}} $\, \cset^p_{l}$ 
by $\cset^{pG}_{l}$, 
$\cset^{pA}_{l}$ by $\cset_{l}^{pGA}$,  $\vm_{\Delta}$ by 
$\vm_{\frac{c}{24}}^G$, and
$\bsvm_{\frac{c}{24}}$ by $\bsvm_{\frac{c}{24}}^G$, then we obtain the 
corresponding definition for 
a $G$-closed adapted ordering 
$\ordering_G$ with ordering kernel $\cset^{pGK}_{l}$. For both cases, 
$\ordering$ and $\ordering_G$, if we replace the annihilation 
conditions by ${\cal K}=\tw^+\oplus\{G_0\}$ we obtain the corresponding
definition for $G$-closed vectors at level $l$, parity $p$. 

Our aim is to introduce an adapted ordering on $\cset^p_{l}$ such that the
resulting ordering kernels are as small as possible. There are two
main reasons to proceed in this way. First, for a given singular 
vector $\Psi^p_{l}$ of $\vm_{\Delta}$ the coefficients of the terms 
corresponding to the ordering kernel $\cset^{pK}_{l}$ 
determine the singular vector uniquely, i.e. if the normal forms of 
two singular vectors agree on
the ordering kernel then the singular vectors are identical. 
This also implies that
for a given singular vector $\Psi^p_{l}$ at least one element of 
the adapted kernel
must be a non-trivial term of $\Psi^p_{l}$, otherwise $\Psi^p_{l}$ 
would agree on the whole adapted kernel 
with the trivial vector and would therefore be trivial itself.
Secondly, the number of elements in the ordering kernel $\cset^{pK}_{l}$
is an upper limit for the singular dimensions of
$\vm_{\Delta}$ at level $l$, parity $p$. These results are the content 
of two theorems which we proved 
in a very general setting in Ref. \icite{p6sdim1}. Due to their importance 
for our later considerations we repeat these theorems here adapted to 
the particular case of the twisted $N=2$ algebra $\tw$.

\bth \label{th:kernel}
Let $\ordering$ denote an adapted ordering on $\cset^{pA}_{l}$ 
at level $l$, parity $p$, with ordering kernel $\cset^{pK}_{l}$ 
for a given Verma module $\vm_{\Delta}$ and annihilation operators 
${\cal K}=\tw^+$. If the normal form of two vectors $\Psi^{p,1}_{l}$
and $\Psi^{p,2}_{l}$ at the same level $l$ and parity
$p$, both satisfying the highest weight conditions, 
have $c_{X}^1=c_{X}^2$ for all terms $X\in\cset^{pK}_{l}$, then
\bea
\Psi^{p,1}_{l} &\equiv & \Psi^{p,2}_{l}\,.
\eea
\eth

\bth \label{th:dims}
Let $\ordering$ denote an adapted ordering on $\cset^{pA}_{l}$ 
at level $l$, parity $p$, with ordering kernel $\cset^{pK}_{l}$ 
for a given Verma module $\vm_{\Delta}$ and annihilation
operators ${\cal K}=\tw^+$.
If the ordering kernel $\cset^{pK}_{l}$ has $n$ elements,
then there are at most $n$ linearly independent singular vectors
$\Psi^p_{l}$ in $\vm_{\Delta}$ at level $l$ with parity $p$.
\eth

Again, we can replace $\ordering$ on $\cset^p_{l}$ by $\ordering_G$ on 
$\cset^{pG}_{l}$
in order to obtain both theorems for $G$-closed Verma modules 
$\vm^G_{\frac{c}{24}}$.
If we extend the annihilation operators from ${\cal K}=\tw^+$ 
to ${\cal K}=\tw^+\oplus\{G_0\}$, which
normally leads to a smaller ordering kernel, then
both theorems hold with this smaller ordering kernel for $G$-closed 
singular vectors $\Psi^{pG}_{l}$.

%%%%%%%%%%%%%%%%%%%%%%%%%%%%%%%%%%%%%%%%%%%%%%%%%%%%%%%%%%%
%                                                         %
%          adapted orderings for the twisted N=2          %
%                                                         %
%%%%%%%%%%%%%%%%%%%%%%%%%%%%%%%%%%%%%%%%%%%%%%%%%%%%%%%%%%%

\section{Adapted ordering for the twisted $N=2$ algebra}
\label{sec:adapted}

In this section we will define an adapted ordering on $\cset^p_{l}$, 
given by \eq{\ref{eq:cset}},
and compute its ordering kernel which will later 
turn out to be the smallest possible ordering kernel. We will conclude 
this section by transferring these results to the $G$-closed case
$\cset^{pG}_{l}$. For convenience, we shall first give an ordering on the 
sets $\lset_m$, $\tset_s$, and $\gset_k$.

\bdf \label{def:lorder}
Let $Y$ denote either $\lset$, $\tset$, or $\gset$,
(but the same throughout this definition) and take two elements $Y_i$,
$i=1,2$, such that $Y_i=Z^i_{-k_{\length{Y_i}}^i}
\ldots Z^i_{-k^i_1}$, $\levelt{Y_i}=k^i$ for $k^i\in\bbbnuo$, or
$Y_i=1$, $i=1,2$, with $Z^i_{-k^i_j}$ being generators of the type
$L_{-k^i_j}$, $T_{-k^i_j}$, or $G_{-k^i_j}$ depending on 
whether $Y$ denotes $\lset$, $\tset$, or $\gset$ respectively.
For $Y_1\neq Y_2$ 
we compute the index\footnote{For subsets of $\bbbn$ we define 
$\min\emptyset=0$.} $j_0=\min\{j:k^1_j-k^2_j\neq 0,
j=1,\ldots,\min(\length{Y_1},\length{Y_2})\}$. If non-trivial,
$j_0$ is the index for which the levels of the generators 
in $Y_1$ and $Y_2$ first disagree when read from the right to the left. 
For $j_0>0$ we then define
\bea
Y_1\osm{Y}Y_2 & {\rm if} & k^1_{j_0}<k^2_{j_0} \,.
\eea
If, however, $j_0=0$, we set
\bea
Y_1\osm{Y}Y_2 & {\rm if} & \length{Y_1}>\length{Y_2} \,.
% shorter sequences are defined to be larger !!!!
\eea
\edf

\noi
In order to get more familiar with these definitions let us consider 
the following examples: 
\bea
L_{-2}L_{-2}L_{-1} & \osm{L} & L_{-5}L_{-1} \;\com \ \ {\rm where} \;\;
j_0=2 \com \nn \\
G_{-3}G_{-\frac{5}{2}}G_{-1} & \osm{G} & G_{-\frac{5}{2}}G_{-1}
\;\com \ \ {\rm where} \;\; j_0=0 \com  \nn \\
T_{-\frac{3}{2}} & \osm{T} & 1 \;\com \ \ {\rm where} \;\; j_0=0 \pkt
\eea

\noi 
We can now define an ordering on $\cset^p_{l}$ which 
will turn out to be adapted with a sufficiently small ordering kernel.
\bdf \label{def:cord}
On the set $\cset^p_{l}$, $l\in\bbbnu$, $p\in\{\pm\}$ we introduce the 
total ordering $\ordering$. 
For two elements $X_1,X_2\in\cset^p_{l}$, $X_1\neq X_2$
with $X_i=L^i T^i G^i T_{-\frac{1}{2}}^{n^i}G_{-\frac{1}{2}}^{r^i_1}
G_{0}^{r^i_2}$, $L^i\in\lset_{m_i}$,
$T^i\in\tset_{s_i}$, $G^i\in\gset_{k_i}$ for
some $m_i,n^i\in\bbbno$, $s_i,k_i\in\bbbnu$ , $\, r_1^i,r_2^i\in\{0,1\}$, 
$i=1,2$, we define
\bea
X_1\osm{\ordering}X_2 & {\rm if} & n^1>n^2 \,. \label{eq:cord1}
\eea
For $n^1=n^2$ we set
\bea
X_1\osm{\ordering}X_2 & {\rm if} & r^1_1>r^2_1 \,.  \label{eq:cord2}
\eea
If $r^1_1=r^2_1$ then we set
\bea
X_1\osm{\ordering}X_2 & {\rm if} & G^1\osm{G} G^2 \,. \label{eq:cord3}
\eea
In the case that $G^1=G^2$ we then define
\bea
X_1\osm{\ordering}X_2 & {\rm if} & L^1\osm{L} L^2 \,. \label{eq:cord4}
\eea
If further $L^1=L^2$ we set
\bea
X_1\osm{\ordering}X_2 & {\rm if} & T^1\osm{T} T^2 \,, \label{eq:cord5}
\eea
which finally has to give an answer. For $X_1=X_2$ we define
$\, X_1\osm{\ordering}X_2$ and $\, X_2\osm{\ordering}X_1$.
\edf 

Definition \refoth{\ref{def:cord}} is well-defined since one obtains
an answer for any pair $X_1, X_2\in\cset^p_{l}$, $X_1\neq X_2$ 
after going through \eqs{\ref{eq:cord1}}-\eqoth{\ref{eq:cord5}},
and hence the ordering $\ordering$ proves to be a total ordering 
on $\cset^p_{l}$. Namely, if \eqs{\ref{eq:cord1}}-\eqoth{\ref{eq:cord5}}
do not give an answer on the ordering of $X_1$ and $X_2$, then 
obviously $X_1$ and $X_2$ are of the form 
$X_i=LTGT_{-\frac{1}{2}}^nG_{-\half}^{r_1}G_{0}^{r_2^i}$,
with common $L$, $T$, $G$, $n$, and also $r_1$.
The fact that both $X_1$ and $X_2$ have the same parity $p$ implies
further $r_2^1=r_2^2$ and ultimately $X_1=X_2$. 

It is easy to see that the $\ordering$-smallest element of $\cset^+_{l}$ 
is $T_{-\frac{1}{2}}^{2l}$ followed by 
$T_{-\frac{1}{2}}^{2l-1}G_{-\frac{1}{2}}G_0$ whilst 
the $\ordering$-smallest element of $\cset^-_{l}$
is $T_{-\frac{1}{2}}^{2l}G_{0}$ followed by 
$T_{-\frac{1}{2}}^{2l-1}G_{-\frac{1}{2}}$.
We will now show that the ordering $\ordering$
is adapted and we will compute the ordering kernels. It will turn out
that the elements of the ordering kernels are exactly the four elements 
just mentioned and therefore they identify all singular vectors 
according to theorem \refoth{\ref{th:kernel}}.
In section \refoth{\ref{sec:examples}} 
we shall give examples of singular vectors at low levels which
show that these are indeed the smallest ordering 
kernels for general values of $\Delta$.

\bth \label{th:adkernel}
For all central terms $c\in\bbbc$, 
the ordering $\ordering$
is adapted to $\cset^p_{l}$ for all
Verma modules $\vm_{\Delta}$, for all levels
$l$ with $l\in\bbbnuo$, and for both parities $p\in\{\pm\}$. 
The ordering kernels are given by the following
tables for all levels\footnote{Note that for level $l=0$ 
some of the kernel elements obviously do not exist.} 
$l$.

\btab{|r|l|}
\hline \label{tab:adkern1}
 $(-1)^F$ & ordering kernel \\
\hline
$+1$ &  $\{T_{-\frac{1}{2}}^{2l}, T_{-\frac{1}{2}}^{2l-1}
G_{-\frac{1}{2}}G_0\}$ \\
\hline
$-1$ &   $\{T_{-\frac{1}{2}}^{2l}G_0, T_{-\frac{1}{2}}^{2l-1}
G_{-\frac{1}{2}}\}$ \\
\hline
\etab{Ordering kernels for $\ordering$,
annihilation operators $\tw^+$.}

\btab{|r|l|}
\hline \label{tab:adkern2}
 $(-1)^F$ & ordering kernel \\
\hline
$+1$ &  $\{T_{-\frac{1}{2}}^{2l}\}$ \\
\hline
$-1$ &   $\{T_{-\frac{1}{2}}^{2l}G_0\}$ \\
\hline
\etab{Ordering kernels for $\ordering$,
annihilation operators $\tw^+$ and $G_0$.}
\eth

For the proof of theorem \refoth{\ref{th:adkernel}}
we follow the lines of the proof for the topological $N=2$
algebra given in Ref. \icite{p6sdim1}. There, however, our strategy for 
the proof 
was to find annihilation operators that were able to produce an additional
operator $L_{-1}$ which would create a term that cannot be obtained by any
$\ordering$-bigger term. Obviously, in our present case $T_{-\frac{1}{2}}$ 
has taken over the r\^ole that was played by $L_{-1}$ in the topological 
$N=2$ case. (As a matter of fact this is the first time that
this r\^ole is played by a generator different from 
$L_{-1}$). The following proof makes clear 
that in this case the adapted ordering method works for powers 
of $T_{-\frac{1}{2}}$.

\bprf
Let us start with the most general term $X_0$ at level $l$, parity $p$,
in $\cset^p_{l}$, $l\in\bbbnuo$, $p\in\{\pm\}$.
\bea
X_0 &=& L^0 T^0 G^0 T_{-\half}^{n} G_{-\half}^{r_1} G_0^{r_2} \; 
\in\cset^{p}_{l} \com \label{eq:x0}
\eea
with 
\bea
L^0 & = & L_{-m_{\length{L^0}}}\ldots L_{-m_1}\in\lset_m \,, \nn \\
T^0 & = & T_{-s_{\length{T^0}}}\ldots T_{-s_1}\in\tset_s \,, \nn \\
G^0 & = & G_{-k_{\length{G^0}}}\ldots G_{-k_1}\in\gset_k \,,
\eea
$m,n\in\bbbno$, $s,k\in\bbbnuo$, $r_1,r_2\in\{0,1\}$, such that
$l=m+s+k+\frac{n}{2}+\frac{r_1}{2}$ and $p=(-1)^{\length{G^0}+r_1+r_2}$. 
We then construct
the vector $\Psi^0=X_0\ket{\Delta}$.

In the case $G^0\not =1$ we look at $G_{-k_1}$ which necessarily has
$k_1>\half$ and consider
the positive operator $G_{k_1-\half}$. The commutation relations of
$G_{-k_1}$ and $G_{k_1-\half}$ create a generator $T_{-\half}$ (note that
$2k_1-\half>0$). Therefore, acting with the positive operator 
$G_{k_1-\half}$
on $\Psi_0=X_0\ket{\Delta}$ we obtain a non-trivial term $X^G$ 
in $G_{k_1-\half}\Psi^0$ of the form
\bea
X^G &=& L^0 T^0 \tilde{G}^0 T_{-\half}^{n+1} G_{-\half}^{r_1}G_0^{r_2} 
\;\;\;\in\cset^{-p}_{l-k_1+\half} \com \nn \\
\tilde{G}^0 &=& G_{-k_{\length{G^0}}}\ldots G_{-k_2} \com
\eea
or $\tilde{G}^0=1$ in the case $\length{G^0}=1$. Any other term 
$Y\in\cset^{p}_{l}$ which also produces $X^G$ under the action of 
$G_{k_1-\half}$ also creates at least one generator $T_{-\half}$ under
the action of $G_{k_1-\half}$ or is already $\ordering$-smaller 
than $X_0$ since $Y$ would already have 
more generators $T_{-\half}$ than $X_0$. The latter case is irrelevant
for us in view of definition \refoth{\ref{def:adapt}} on adapted orderings.
The action of one positive operator can however create at most
one new generator. We can therefore restrict ourselves to terms 
$Y$ of the form
\bea
Y &=& L^Y T^Y G^Y T_{-\half}^{n} G_{-\half}^{r^Y_1} G_0^{r^Y_2} \; 
\in\cset^{p}_{l} \com
\eea
with the same number $n$ of generators $T_{-\half}$ than $X_0$. 
If $G_{k_1-\half}$ creates $X^G$ acting on $Y$, 
then the additional $T_{-\frac{1}{2}}$ could only be created in three ways.
First, $G_{k_1-\half}$ commutes with generators in $G^Y$ and creates 
$T_{-\half}$. Second, $G_{k_1-\half}$ commutes with generators 
in $L^YT^Y$ and creates $G_{k^{\prime}}$ with
$k^{\prime}<k_1$ which furthermore commutes with generators in $G^Y$ to
create $T_{-\half}$. Third, $G_{k_1-\half}$ commutes with generators 
in $L^YT^YG^Y$ and creates $G_0$ which then commutes with 
$G_{-\half}$ in the case of $r^Y_1=1$ to create
$T_{-\half}$. In the third case $r^Y_1=1$ would obviously lead to $Y$
being $\ordering$-smaller than $X_0$ because then $r_1=0$ (as
$X^G$ is the term produced from $Y$ in this way), and due to 
\eq{\ref{eq:cord2}}. In the first and
second cases we obviously need to have a generator $G_{-k}$ in $G^Y$ with
$\half<k<k_1$, or $G_{-k_1}$ itself is contained in $G^Y$ and 
$G_{k_1-\half}$ commutes only with this generator. In the latter case 
$X_0$ would consequently equal $Y$ and in the former case $Y$ is
again $\ordering$-smaller than $X_0$ according to \eq{\ref{eq:cord3}}. 
Hence, we have shown that the ordering $\ordering$ is adapted on the set 
of terms $X_0$ of the form \eq{\ref{eq:x0}} with $G^0\not =1$. 

Let us therefore continue with terms $X_0$ of the form
\bea
X_0 &=& L^0 T^0 T_{-\half}^{n} G_{-\half}^{r_1} G_0^{r_2} \; 
\in\cset^{p}_{l} \pkt \label{eq:x03}
\eea
If $L^0\not =1$ we consider the positive operator $T_{m_1-\half}$. Acting 
with $T_{m_1-\half}$ on $\Psi_0=X_0\ket{\Delta}$ creates a non-trivial term 
\bea
X^T &=& \tilde{L}^0 T^0 T_{-\half}^{n+1} G_{-\half}^{r_1} G_0^{r_2} 
\;\;\;\in\cset^{p}_{l-m_1+\half} \com \nn \\
\tilde{L}^0 &=& L_{-m_{\length{L^0}}}\ldots L_{-m_2} \com
\eea
or $\tilde{L}=1$ in the case $\length{L^0}=1$. Again, we find that any term 
$Y\in\cset^{p}_{l}$ which is not $\ordering$-smaller than $X_0$ but still 
produces $X^T$ under the action of $T_{m_1-\half}$ needs to create 
exactly one generator $T_{-\half}$ and thus $n^Y=n$. As in the case before 
we find $r^Y_1=r_1$, and also $G^Y=1$, otherwise $Y\osm{\ordering}X_0$.
We can hence concentrate on terms $Y$ of the form
\bea
Y &=& L^Y T^Y T_{-\half}^{n} G_{-\half}^{r_1} G_0^{r_2} \; 
\in\cset^{p}_{l} \com
\eea
where also $r^Y_2=r_2$ due to parity equality with $X_0$. The only way of
creating $T_{-\half}$ from $Y$ under the action of $T_{m_1-\half}$ 
is via commutation with
generators in $L^Y$ as the commutation with generators in $T^Y$ 
does not produce any new generators. Thus, 
$T_{m_1-\half}$ can produce $T_{-\half}$ in one step by commuting
with $L_{-m_1}$ if this is contained in $L^Y$. In this case, 
however, $Y$ turns out to be equal to $X_0$. Another possibility is that
$T_{m_1-\half}$ produces $T_{-\half}$ in more than one step by 
commuting with generators in $L^Y$ which obviously requires $L^Y$ 
to contain generators $L_{-m}$ with $1\leq m<m_1$ and therefore 
$Y\osm{\ordering}X_0$ due to \eq{\ref{eq:cord4}}. We therefore have 
shown that the ordering $\ordering$ is adapted on the set 
of terms $X_0$ of the form \eq{\ref{eq:x0}} with
$G^0\not =1$ or $L^0\not =1$. 

For the next step we can thus assume that $X_0$ is of the form
\bea
X_0 &=& T^0 T_{-\half}^{n} G_{-\half}^{r_1} G_0^{r_2} \; 
\in\cset^{p}_{l} \pkt \label{eq:x04}
\eea
If $T^0\not =1$ we take the positive operator $L_{s_1-\half}$. Acting with 
$L_{s_1-\half}$ on $\Psi_0=X_0\ket{\Delta}$ creates a non-trivial term
\bea
X^L &=& \tilde{T}^0 T_{-\half}^{n+1} G_{-\half}^{r_1} G_0^{r_2} 
\;\;\;\in\cset^{p}_{l-s_1+\half} \com \nn \\
\tilde{T}^0 &=& T_{-s_{\length{T^0}}}\ldots T_{-s_2} \com
\eea
or $\tilde{T}=1$ in the case $\length{T^0}=1$. Similar arguments as 
above allow us to restrict ourselves to terms $Y$ of the form
\bea
Y &=& T^Y T_{-\half}^{n} G_{-\half}^{r_1} G_0^{r_2} \; \in\cset^{p}_{l} \com
\eea
as all other terms $Y$ that also produce $X^L$ under the action of 
$L_{s_1-\half}$ would automatically be $\ordering$-smaller than $X_0$.
Similarly as before, the only way of creating $T_{-\half}$ from $Y$ 
under the action of $L_{s_1-\half}$ is by commuting $L_{s_1-\half}$ 
with generators in $T^Y$. Exactly the same arguments as
for $L^Y$ together with \eq{\ref{eq:cord5}} show that either $Y=X_0$ 
or $Y\osm{\ordering}X_0$.
Hence, the ordering $\ordering$ is adapted on the set of terms $X_0$ 
of the form \eq{\ref{eq:x0}} with
$G^0\not =1$ or $L^0\not =1$ or $T^0\not = 1$. This already proves 
the ordering kernels
of table \tab{\ref{tab:adkern1}}.

Finally, if we also have $G_0$ as annihilation operator, we can then 
act with $G_0$ on terms $X_0$ of the form
\bea
X_0 &=& T_{-\half}^{n} G_{-\half} G_0^{r_2} \; \in\cset^{p}_{l} \com 
\label{eq:x05}
\eea
which creates a non-trivial term
\bea
X^{\prime} &=& T_{-\half}^{n+1} G_0^{r_2} \;\;\;\in\cset^{-p}_{l} \pkt 
\eea
Obviously, the only way of creating $T_{-\half}$ by commuting 
$G_0$ with products of generators
is by commuting it with $G_{-\half}$. Therefore any other 
term $Y$ that creates $X^{\prime}$
under the action of $G_0$ and is not $\ordering$-smaller 
than $X_0$ would also need to create one
$T_{-\half}$ and thus also have $r^Y_1=r_1=1$, besides
$n^Y=n$. Trivially, we find in this case that $Y=X_0$. 
This proves the ordering kernel of table \tab{\ref{tab:adkern2}} 
which finally completes our proof of theorem \refoth{\ref{th:adkernel}}.
\eprf

To conclude this subsection we mention that we can easily
transfer these results to $G$-closed Verma modules.
It is fairly straightforward to see that the above proof also 
holds if we set everywhere $r_2\equiv 0$.
This however leads exactly to the $G$-closed case. Therefore, 
the same ordering $\ordering$
of definition \refoth{\ref{def:cord}} will also serve as ordering 
$\ordering_G$ on $\cset^{pG}_{l}$ in
the case of a $G$-closed Verma module $\vm_{\frac{c}{24}}^G$. Setting
$r_2\equiv 0$ everywhere in the proof of theorem 
\refoth{\ref{th:adkernel}} obviously reduces the ordering kernels. 
One easily finds the following theorem.
\bth \label{th:adkernel2}
For all central terms $c\in\bbbc$, 
the ordering $\ordering$
is adapted to $\cset^{pG}_{l}$ for all
Verma modules $\vm^G_{\frac{c}{24}}$, for all levels
$l$ with $l\in\bbbnuo$, and for both parities $p\in\{\pm\}$. 
Ordering kernels are given by the following
tables for all levels $l$.

\btab{|r|l|}
\hline \label{tab:adkern3}
 $(-1)^F$ & ordering kernel \\
\hline
$+1$ &  $\{T_{-\frac{1}{2}}^{2l}\}$ \\
\hline
$-1$ &   $\{T_{-\frac{1}{2}}^{2l-1}G_{-\frac{1}{2}}\}$ \\
\hline
\etab{Ordering kernels for $\ordering$ on $\cset^{pG}_{l}$,
annihilation operators $\tw^+$.}

%\btab{|r|l|}
%\hline \label{tab:adkern4}
% $(-1)^F$ & ordering kernel \\
%\hline
%$+1$ &  $\{T_{-\frac{1}{2}}^{2l}\}$ \\
%\hline
%$-1$ &   $\{\}$ \\
%\hline
%\etab{Ordering kernels for $\ordering$ on $\cset^{pG}_{l}$,,
%annihilation operators $\tw^+$ and $G_0$.}
\eth

%%%%%%%%%%%%%%%%%%%%%%%%%%%%%%%%%%%%%%%%%%%%%%%%%%%%%%%%%%%
%                                                         %
%        singular dimensions of the twisted N=2           %
%                                                         %
%%%%%%%%%%%%%%%%%%%%%%%%%%%%%%%%%%%%%%%%%%%%%%%%%%%%%%%%%%%

\section{Singular dimensions and singular vectors of the twisted $N=2$ 
algebra}
\label{sec:sdim}
As a consequence of the results of the previous section we will now 
be able to give an upper limit for
the dimensions of a singular vector space at a given level $l\in\bbbnuo$ 
with fixed parity $p\in\{\pm\}$. 
In addition, we will use the ordering kernels 
in order to uniquely identify all singular vectors (primitive as well as
secondary) of the twisted $N=2$ superconformal algebra.
In section \refoth{\ref{sec:examples}} we shall 
give explicit examples
which prove that for particular Verma modules the upper limits are 
indeed the maximal singular dimensions. 

Theorem \refoth{\ref{th:dims}} tells us that an upper limit for the 
dimension of a
space of singular vectors at fixed level $l$ and parity $p$ is given 
by the number of elements in the corresponding 
ordering kernel $\cset^{pK}_{l}$. Using the tables of ordering kernels
of theorem \refoth{\ref{th:adkernel}} and theorem 
\refoth{\ref{th:adkernel2}} we can easily give the following
maximal singular dimensions.

\bth
For singular vectors $\Psi^p_{l}$ and $G$-closed singular vectors
$\Psi^{pG}_{l}$ in the Verma modules   
$\vm_{\Delta}$ and $\vm_{\frac{c}{24}}^G$ (both), one finds the
following upper limits for the number of linearly independent
singular vectors at the same level $l\in\bbbnuo$ and with the same 
parity $p\in\{\pm\}$.
\btab{|l|c|c|}
\hline \label{tab:dim1}
 & $p=+1$ & $p=-1$ \\
\hline 
$\Psi^p_{l}\in\vm_{\Delta}$ & 
$2$ & $2$ \\
\hline 
$\Psi^p_{l}\in\vm_{\frac{c}{24}}^G$ & 
$1$ & $1$ \\
\hline 
$\Psi^{pG}_{\frac{c}{24}-\Delta}\in\vm_{\Delta}$ & 
$1$ & $1$ \\
\hline 
$\Psi^{pG}_{0}\in\vm_{\frac{c}{24}}^G$ & 
$0$ & $0$ \\
\hline
\etab{Maximal dimensions for singular spaces of the twisted 
$N=2$ superconformal algebra.}
\eth
				      
Surely, not at each level the singular spaces will have the 
dimensions given in the previous tables. In fact, for 
most levels in most Verma modules there will not be any singular vectors
at all. The determinant formulae of the inner product matrix\cite{bfk}, 
given in section \refoth{\ref{sec:alg}}, indicate in which Verma modules 
and at which levels singular vectors appear (for the first time). In the 
following section we shall give examples of
these singular vectors for low levels. From theorem 
\refoth{\ref{th:kernel}} we know that
singular vectors can only exist if they
contain in their normal form at least one non-trivial term of the 
corresponding ordering kernel of theorem \refoth{\ref{th:adkernel}} or 
\refoth{\ref{th:adkernel2}}. Furthermore, if the normal forms of 
two singular vectors agree
in the coefficients of the terms of the ordering kernel, 
then the singular vectors are identical.
This allows us to identify existing singular vectors simply by 
their coefficients with respect
to the terms of the ordering kernel. For the cases of  
\tab{\ref{tab:dim1}} with
singular dimension $1$ this implies that two singular vectors are
already proven to be proportional if we
can simply show that they are on the same level with the same parity. 
In addition, if the coefficient
of the term of the ordering kernel vanishes, then the singular vector 
is trivial. Thus, the coefficient of $T_{-\half}^{2l}$ 
is already sufficient in order to see if
a singular vector is trivial, in the case of $p=+$ (see tables 
\tab{\ref{tab:adkern2}} and \tab{\ref{tab:adkern3}}), whereas the 
coefficient of $T_{-\half}^{2l}G_0$ or $T_{-\half}^{2l-1}G_{-\half}$
is sufficient in order to decide if a singular vector is trivial 
in the case $p=-$. When the ordering kernel is two-dimensional we need the
coefficients of the two terms in order to uniquely define a singular 
vector. This justifies to introduce an appropriate notation, as follows.

\bdf  \label{def:ab}
Let $\Psi^{\pm}_{l}=\theta^{\pm}_{l}\ket{\Delta}$ be a 
singular vector in $\vm_{\Delta}$ at level
$l\in\bbbnuo$ with parity $\{\pm\}$.  The normal form of $\Psi^+_{l}$
is completely determined by the coefficients of $T_{-\half}^{2l}$ and 
$T_{-\half}^{2l-1}G_{-\half}G_0$,
whilst  the normal form of $\Psi^-_{l}$
is determined by the coefficients of $T_{-\half}^{2l}G_0$ and
$T_{-\half}^{2l-1}G_{-\half}$. 
We thus use the following notation
\bea
(a,b)_{l}^+=\theta^+_{l}=aT_{-\half}^{2l}   + 
bT_{-\half}^{2l-1}G_{-\half}G_0 + \ldots  \com \\
(a,b)_{l}^-=\theta^-_{l}= aT_{-\half}^{2l}G_0   + 
bT_{-\half}^{2l-1}G_{-\half} + \ldots  \com
\eea
for the singular vector operators $\theta^{\pm}_{l}$ given in 
their normal form, where $a,b\in\bbbc$. 
\edf
Notice that definition \refoth{\ref{def:ab}} only makes sense if 
$\Psi^{\pm}_{l}$ is a singular vector. Certainly for most 
pairs $a,b\in\bbbc$ singular vectors do not exist. 

The advantage of this notation lies in the fact that we can easily obtain 
multiplication rules for singular vector operators and construct
descendant singular vectors. (In the context of embedding diagrams, this
issue will be analysed in more detail in a forthcoming publication). For
example, the product of the first terms of the singular vector operators  
\bea
\lefteqn{(a_1T_{-\half}^{2l_1} + b_1T_{-\half}^{2l_1-1}G_{-\half}G_0+\ldots)  
(a_2T_{-\half}^{2l_2} + b_2T_{-\half}^{2l_2-1}G_{-\half}G_0+\ldots)  
\;\;=} \nn \\
&& a_1a_2T_{-\half}^{2(l_1+l_2)}   + (a_1b_2+a_2b_1
-\frac{b_1b_2}{2}) T_{-\half}^{2(l_1+l_2)-1}G_{-\half}G_0+\ldots  \com
\eea
gives us the multiplication rule for $(a_1,b_1)_{l_1}^+(a_2,b_2)_{l_2}^+$ 
provided the conformal weights match: $\Delta_2+l_2=\Delta_1$. In the same
way we can easily find similar rules for products involving negative 
parity operators. The following theorem summarizes these results.
\bth
Given two singular vector operators $\theta^{p_1}_{l_1}$ and 
$\theta^{p_2}_{l_2}$
for the Verma modules $\vm_{\Delta_1}$ and $\vm_{\Delta_2}$ with 
$\Delta_2+l_2=\Delta_1$ then
$\theta^{p_1}_{l_1}\theta^{p_2}_{l_2}\ket{\Delta_2}$ is either trivial 
or singular in $\vm_{\Delta_2}$
at level $l_1+l_2$ with parity $p_1p_2$. The resulting singular vector 
can be expressed in the following way, depending on the multiplication
rules which in turn depend on the parities:
\bea
(a_1,b_1)^+_{l_1} \, (a_2,b_2)_{l_2}^+ \ket{\Delta_2} &=& 
(a_1a_2, \, a_1b_2+a_2b_1-\frac{b_1b_2}{2})^+_{l_1+l_2} 
\ket{\Delta_2} \,, \\
(a_1,b_1)^+_{l_1} \, (a_2,b_2)_{l_2}^- \ket{\Delta_2} &=& 
(a_1a_2, \, a_1b_2+a_2b_1[\Delta_2-\frac{c}{24}] 
-\frac{b_1b_2}{2})^-_{l_1+l_2} 
\ket{\Delta_2} \,, \\
(a_1,b_1)^-_{l_1} \, (a_2,b_2)_{l_2}^+ \ket{\Delta_2} &=& 
(a_1a_2-\frac{a_1b_2}{2},-2l_2a_1b_2+a_2b_1
-a_1b_2[\Delta_2-\frac{c}{24}])^-_{l_1+l_2} 
\ket{\Delta_2} \,, \\
(a_1,b_1)^-_{l_1} \, (a_2,b_2)_{l_2}^- \ket{\Delta_2} &=& 
(a_1a_2[\Delta_2-\frac{c}{24}]-\frac{a_1b_2}{2}, \,
a_2b_1-2l_2a_1b_2-a_1b_2)^+_{l_1+l_2} 
\ket{\Delta_2} \,.
\eea
\eth

For the vectors $\Psi^{\pm}_{G,l}$ and ${}_{G}\Psi^{\pm}_{l}$, introduced
in \eqs{\ref{eq:psiG1}}-\eqoth{\ref{eq:psiG4}}, we can derive the 
corresponding expressions by looking at operator products of the type
\bea
(aT_{-\half}^{2l}   + bT_{-\half}^{2l-1}G_{-\half}G_0+\ldots)  G_0  &=& 
a T_{-\half}^{2l} G_0   
+ bT_{-\half}^{2l-1}G_{-\half}(L_0-\frac{c}{24})+\ldots \,,
\eea 
and similar products for negative parity vectors. We obtain 
the following results
for $\Psi^{\pm}_{G,l}=(a,b)^{\pm}_{G,l}\ket{\Delta}$ and 
${}_{G}\Psi^{\pm}_{l}={}_G(a,b)^{\pm}_{l} \ket{\Delta}$ :
\bea
{}_G(a,b)^-_l \ket{\Delta} &=& (a-\frac{b}{2},
-b[\Delta-\frac{c}{24}]-2la)^-_l \ket{\Delta} \com \label{eq:thetaG1} \\
(a,b)^-_{G,l} \ket{\Delta} &=& (a,b[\Delta-\frac{c}{24}])^-_l 
\ket{\Delta} \com \\
{}_G(a,b)^+_l \ket{\Delta} &=& (a[\Delta-\frac{c}{24}]
-\frac{b}{2},-2la-b)^+_l \ket{\Delta} \com \\
(a,b)^+_{G,l} \ket{\Delta} &=& (a[\Delta-\frac{c}{24}],b)^+_l 
\ket{\Delta} \pkt \label{eq:thetaG4}
\eea

%%%%%%%%%%%%%%%%%%%%%%%%%%%%%%%%%%%%%%%%%%%%%%%%%%%%%%%%%%%
%                                                         %
%                    Examples                             %
%                                                         %
%%%%%%%%%%%%%%%%%%%%%%%%%%%%%%%%%%%%%%%%%%%%%%%%%%%%%%%%%%%

\section{Level $1/2$, level $1$, and level $3/2$ singular vectors.}
\label{sec:examples}
Before we proceed in analysing the structure of the singular vectors 
we shall first give all singular vectors at the lower levels 1/2, 1 and
3/2. These examples already prove that the upper limits given 
for the singular dimensions are in fact not only upper limits but 
actually maximal singular dimensions for certain Verma modules. The 
determinant formulae\cite{bfk}, given in section \refoth{\ref{sec:alg}}, 
show the existence of the parity $-$ singular vector $\Psi^{-}_{0}$
at level 0 in the Verma modules $\vm_{c\over24}$, and also show the 
existence of the singular
vectors $\Psi^{\pm}_{r,s}$ at level $\frac{rs}{2}$ with parity 
$\pm$, $r,s\in\bbbn$, $s$ odd,
in the Verma modules $\vm_{\Delta_{r,s}(t)}$ with:
\bea
\Delta_{r,s}(t) &=& -\frac{(rt+s)^2}{8t}+\frac{t}{8}+\frac{1}{8} \com 
\label{eq:deltars} \\
c &=& 3+3t \pkt
\eea
These singular vectors $\Psi^{\pm}_{r,s}$ are generically primitive.
However in the case of intersections, with two or more of them in the
same Verma module, some of these singular vectors are actually secondary.

By explicit computer calculations\cite{maple} one finds the following 
singular vectors (one-parameter families of singular vectors, to be 
precise) for parity $+$. The corresponding results for parity $-$ 
can easily be derived using the multiplication rules
\eqs{\ref{eq:thetaG1}}-\eqoth{\ref{eq:thetaG4}}. These singular
vectors are given in appendix \refoth{\ref{app:a}}.

\noi
Level $\frac{1}{2}$:
\bea
\Psi^+_{1,1}(t) &=& \Bigl\{ (t+1) T_{-\half} +4t G_{-\half}G_0 \Bigr\} 
\ket{-\frac{t+1}{8t}} \pkt \label{sv11+}
\eea 

\noi
Level $1$:
\bea
\Psi^+_{2,1}(t) &=& \Bigl\{ (2t+1) T_{-\half}^2 +8t T_{-\half} 
G_{-\half}G_0 +2t(2t+1) L_{-1}  \nn \\
&& -4t(t+1) G_{-1}G_0 \Bigr\}
\ket{-\frac{3t^2+3t+1}{8t}} \pkt
\eea 

\noi
Level $\frac{3}{2}$ has two different singular vectors:
\bea
\Psi^+_{3,1}(t) &=& \Bigl\{  (3t+1) T_{-\half}^3 +12t T_{-\half}^2G_{-\half}G_0 
+6t(3t+2)L_{-1}T_{-\half} \nn \\
&&   -4t(4t+3) G_{-1}T_{-\half}G_0 + 8t^2 L_{-1}G_{-\half}G_0 + 2t(3t+1) 
G_{-1}G_{-\half}  \nn\\
&& +2t(t-1)(3t+1) T_{-\thalf} +8t(t^2-t-1)  G_{-\thalf}G_0 \Bigr\} 
\ket{-\frac{8t^2+5t+1}{8t}} \com \\
\Psi^+_{1,3}(t) &=&\Bigl\{  (t+3) T_{-\half}^3 +4t T_{-\half}^2G_{-\half}G_0 
-2t(t+3)L_{-1}T_{-\half} \nn \\
&&   -4t G_{-1}T_{-\half}G_0 - 8t^2 L_{-1}G_{-\half}G_0 + 2t(t+3) 
G_{-1}G_{-\half}  \nn\\
&& -4(t+3) T_{-\thalf} -8t G_{-\thalf}G_0 \Bigr\} \ket{-\frac{5t+9}{8t}} \pkt
\label{sv13+}  \eea 

We can use the notation introduced in definition \refoth{\ref{def:ab}} 
of the previous section in order to
identify the given examples of singular vectors in terms of their 
leading coefficients.
\bea
\Psi^+_{1,1}(t) &=& (t+1,4t)^+_{\half} \ket{-\frac{t+1}{8t}}  \com \\
\Psi^+_{2,1}(t) &=& (2t+1,8t)^+_{1} \ket{-\frac{3t^2+3t+1}{8t}}  \com \\
\Psi^+_{3,1}(t) &=& (3t+1,12t)^+_{\thalf} \ket{-\frac{8t^2+5t+1}{8t}} \com \\
\Psi^+_{1,3}(t) &=& (t+3,4t)^+_{\thalf} \ket{-\frac{5t+9}{8t}}  \pkt
\eea

We should note that at least at these levels there are no other 
singular vectors than the ones
predicted by the determinant formula. In principle, the determinant 
formula only proves the existence
of the singular vector at lowest level in a Verma module given by 
\eq{\ref{eq:deltars}}. 
Continuation arguments extend the existence to the cases where the 
curves of \eq{\ref{eq:deltars}} 
intersect. The determinant formula does {\it a priori} not exclude 
the existence
of so-called {\it isolated} singular vectors which are primitive singular
vectors that may appear at higher levels and are always
{\it hidden} behind a singular vector of type $\Psi^{\pm}_{r,s}$. 
As far as we know, such isolated 
singular vectors have never been found for any chiral algebra. 

As these examples are the  
complete set of singular vectors at levels $\half$, $1$, or $\thalf$ with
positive parity, we observe that the singular spaces at level $\half$ 
and level $1$ are one-dimensional. However, at level $\thalf$ 
we find that the conformal weights $\Delta_{3,1}(t)$
and $\Delta_{1,3}(t)$ intersect for $t=\pm 1$. In these cases 
the corresponding singular vectors are
$\Psi^+_{3,1}(1)=(4,12)^+_{\thalf}\ket{-\frac{7}{4}}$ and 
$\Psi^+_{1,3}(1)=(4,4)^+_{\thalf}\ket{-\frac{7}{4}}$ for $t=1$, and
$\Psi^+_{3,1}(-1)=(-2,-12)^+_{\thalf}\ket{\frac{1}{2}}$ and 
$\Psi^+_{1,3}(-1)=(2,-4)^+_{\thalf}\ket{\frac{1}{2}}$ for $t=-1$.
Obviously, for both values of $t$ these singular vectors are linearly 
independent and thus
span a two-dimensional singular vector space. Therefore we also find 
degenerate singular vectors for the twisted $N=2$ superconformal
algebra as already observed for the three isomorphic $N=2$ superconformal
algebras\cite{cmp1,beatriz1,beatriz2}. 

We should stress that in the Virasoro case for a given level
$l\in\bbbno$ we also have different one-parameter families 
of singular vectors $\xi_{p,q}(t)$ that intersect
for certain values of $t$. However, for Virasoro Verma modules 
these singular vectors at such
intersection points are always linearly dependent. This follows 
indirectly from the proof of Feigin and Fuchs\cite{ff1} 
and directly from the results of Kent\cite{kent} (see also our results 
in Ref. \icite{p6sdim1}). The same happens for the Verma modules of
the $N=1$ superconformal algebras.
The degeneracy found for the Neveu-Schwarz $N=2$ 
algebra\cite{cmp1} (and for the topological and Ramond 
$N=2$ algebras\cite{p6sdim1,beatriz2}) is different in the sense that
one single uncharged vector $\Psi^{NS}_{r,s}$ may on its own describe a
two-dimensional space under certain conditions ($+1$ and $-1$  
charged singular vectors intersecting in the same Verma module 
and the uncharged singular vector becoming a {\it secondary} 
singular vector of them). 
At these points $t=t_0\,$, $\Psi_{r,s}^{NS}(t_0)$ vanishes
and the two-dimensional tangent space becomes a degenerate singular space.
The degeneracy found for the twisted $N=2$ algebra is thus of a
different type since the singular vectors that intersect are generically
{\it primitive}, like the vectors $\Psi^+_{3,1}(1)$ and $\Psi^+_{1,3}(1)$,  
and $\Psi^+_{3,1}(-1)$ and $\Psi^+_{1,3}(-1)$ given above. 
It is in fact unique of its kind in the sense that
it has not been observed for any other chiral algebra so far in this form.
In the following section we will conjecture general expressions
for the twisted $N=2$ singular vectors $\Psi_{r,s}^{\pm}(t)$
(for the relevant coefficients 
of the terms corresponding to the ordering kernel) based on explicit 
examples and also theoretical arguments. From these expressions one
deduces that the parametrised vectors $\Psi_{r,s}^{\pm}(t)$ 
never vanish identically unlike the uncharged singular vectors
of the Neveu-Schwarz $N=2$ algebra at the degenerate points.  

%%%%%%%%%%%%%%%%%%%%%%%%%%%%%%%%%%%%%%%%%%%%%%%%%%%%%%%%%%%
%                                                         %
%                   Classification                        %
%                                                         %
%%%%%%%%%%%%%%%%%%%%%%%%%%%%%%%%%%%%%%%%%%%%%%%%%%%%%%%%%%%

\section{All twisted $N=2$ singular vectors.}
\label{sec:classification}
We have learned in section \refoth{\ref{sec:alg}} that every positive parity 
singular vector  $\Psi^+_{r,s}$, corresponding to an operator $(a,b)^+_l$, 
implies the existence of two singular vectors with negative parity 
at the same level $l$, corresponding to the operators ${}_G(a,b)^-_l$ and 
$(a,b)^-_{G,l}$ (and the other way around for $\Psi^-_{r,s}$). 
These two negative (or positive) parity singular vectors 
may or may not be linearly independent. If they were always linearly 
independent, then whenever a singular vector exists we would find a 
two-dimensional singular space with the opposite parity. However, as was
pointed out, the fact that $P(0)=1$ in the determinant formula, given by
\eq{\ref{detform}}, points towards the existence of, generically, one 
singular vector of each parity at the same level in the same Verma module 
$\vm_{\Delta_{r,s}}$. Nevertheless, as we will see , even in the case where 
${}_G(a,b)^-_l$ and $(a,b)^-_{G,l}$ (or ${}_G(a,b)^+_l$ and $(a,b)^+_{G,l}$)
are proportional the maximal singular dimension may still be $2$. 
In order to analyse this case 
in more detail we will use the expressions given by 
\eqs{\ref{eq:thetaG1}}-\eqoth{\ref{eq:thetaG4}} for ${}_G(a,b)^-_{l}$, 
$(a,b)^-_{G,l}\, $, ${}_G(a,b)^+_{l}$ and $(a,b)^+_{G,l}\, $.  

Requiring that the two negative parity singular vectors are proportional, 
and using the explicit expressions for $\Delta_{r,s}(t)$, given by 
\eqs{\ref{eq:deltars}},
leads to two solutions: $\frac{b}{a}=4\frac{rt}{rt+s}$ or 
$\frac{b}{a}=4\frac{s}{rt+s}$.
Note that under the proportionality assumption, the case $a=0$ also 
implies $b=0$ and thus $(a,b)^+_l\equiv 0$,
which we want to exclude. Therefore, if either of the two 
conditions on the ratio $\frac{b}{a}$ is
satisfied, then the two negative parity singular vectors 
derived from a positive parity singular
vector would be proportional and thus the corresponding singular dimension 
may well only be $1$. Surprisingly enough, all examples found so far 
satisfy always the first of these two conditions. These examples 
include the singular vectors at levels $\half$, $1$ and $\frac{3}{2}$ given
in section \refoth{\ref{sec:examples}} as well as singular vectors at 
levels $2$ and $\frac{5}{2}$. We believe that there must be a deep reason
behind this result, i.e. underlying the fact that all the singular vectors
we have computed satisfy the same condition (out of two possible ones)
that prevents a positive parity singular vector to impose the existence 
of a two-dimensional singular space of negative parity. For this reason 
we conjecture that the positive parity singular vectors 
$\Psi^+_{r,s} = (a,b)^+_{\frac{rs}{2}}\ket{\Delta_{r,s}}$ follow the ratio
$\frac{b}{a}=4\frac{rt}{rt+s}$. From this we can easily derive 
the corresponding ratio $\frac{b'}{a'}$ for the negative parity singular 
vectors $\Psi_{r,s}^- = (a',b')^-_{\frac{rs}{2}}\ket{\Delta_{r,s}}$.
These results allow us to conjecture the following expressions for the
twisted $N=2$ singular vectors. 
\bcj \label{cj:ab}
The singular vectors $\Psi^{\pm}_{r,s}$ in the Verma modules 
$\vm_{\Delta_{r,s}(t)}$ are given by:
\bea
\Psi^+_{r,s}(t) & = & (rt+s,4rt)^+_{\frac{rs}{2}}\ket{\Delta_{r,s}(t)} \com 
\label{eq:psirs1} \\
\Psi^-_{r,s}(t) & = & (-\frac{2}{r}, rt+s)^-_{\frac{rs}{2}}
\ket{\Delta_{r,s}(t)} \pkt \label{eq:psirs2}
\eea
\ecj

Also for the negative parity singular vectors $\Psi_{r,s}^-$,
the last two equations of
\eqs{\ref{eq:thetaG1}}-\eqoth{\ref{eq:thetaG4}} both lead to
$\Psi^+_{r,s}$ rather than to a pair of different singular
vectors. The only exceptional cases occur when $\Psi_{r,s}^{\pm}$ or
$\vm_{\Delta}$ become
$G$-closed and consequently some of these vectors vanish. This will be
considered in the following sections.

Using the superconformal fusion rules it may be possible to derive explicit
expressions for all the coefficients of the singular vectors 
\eqs{\ref{eq:psirs1}}-\eqoth{\ref{eq:psirs2}} 
in the spirit of Bauer, di Francesco, Itzykson, and 
Zuber\cite{bauer1,bauer2},
as it was possible for the Neveu-Schwarz $N=2$ case\cite{ijmpa1}. 
This could also lead to a proof of conjecture \refoth{\ref{cj:ab}}.
For this purpose
it would be necessary to develop\cite{p11vanish} first a twisted 
superfield formalism following the approach
of Ref. \icite{rmp1} for the Neveu-Schwarz algebras. 

Based on conjecture \refoth{\ref{cj:ab}} it is easy to compute 
the Verma modules with degenerate singular vectors, i.e. with singular 
spaces of dimension $2$. For a given level $l\in\bbbnuo$ there are
normally multiple solutions to the number theoretical problem $rs=2l$, 
$r,s\in\bbbn$, s odd. It happens the first time for level
$l=\frac{3}{2}$ that we find multiple solutions $r=1$, $s=3$ and $r=3$, 
$s=1$. In the general case one factorises the integer $2l$ in its prime 
factors  
\bea
2l &=& 2^n \prod_{i=1}^{P} p_i^{n_i} \com
\eea
with $p_i>2$ distinct primes and $n\in\bbbn$, $n_i\in\bbbno$. 
The solutions to $rs=2l$, $r,s\in\bbbn$, s odd are then given by
\bea
r_{\pi} &=& 2^n \prod_{i=1}^P p_i^{k_i} \com \nn \\
s_{\pi} &=& \prod_{i=1}^P p_i^{n_i-k_i} \com \label{eq:rspi}
\eea
for each $P$-tuple $\pi=(k_1,k_2,\ldots,k_P)\in\bbbno^P$, 
with $k_i \leq n_i, \, \forall i=1,\ldots,P $. 
For example, at level $l=630$ we find the prime factorisation 
$2l=2^2 \times 3^2 \times 5 \times 7$.
Hence, we find 12 3-tuples $(k_1,k_2,k_3)\in\bbbno^3$ with 
$k_1\leq 2$ and $k_2,k_3\leq 1$.
We should note that for each level $l\in\bbbnu$ there exists at 
least one solution: 
$r=2l$, $s=1$. The number of solutions to $rs=2l$, $r,s\in\bbbn$, 
s odd is thus determined by the number
of $P$-tuples $\pi=(k_1,\ldots,k_P)$ with $\pi\in\bbbno^P$ and 
$\pi\leq(n_1,\ldots,n_P)$ for the prime factorisation of $2l$. 
Hence, there are\footnote{We define the empty product as 
$\prod_{i=1}^{0}=1$.} $\Pi_l=\prod_{i=1}^P (n_i+1)$ solutions to 
$rs=2l$, $r,s\in\bbbn$, s odd.

Due to our conjecture \refoth{\ref{cj:ab}} we have generically 
only one-dimensional singular spaces defined by the singular 
vectors $\Psi^+_{r,s}$ (or $\Psi^-_{r,s}$) at level $l=\frac{rs}{2}$ for 
all $\Pi_l$ solutions of $r$ and $s$. That is, these singular vectors
with the same parity and level are located in different Verma modules. The
most interesting Verma modules are those, where some of these $\Pi_l$ 
solutions intersect and may hence lead to two-dimensional singular spaces 
provided the corresponding singular vectors are not proportional. 
Using conjecture \refoth{\ref{cj:ab}} 
we shall now investigate such cases of degeneration. Let us note again 
that the singular dimension can not be bigger than $2$ even though a
priori we would expect that even more than $2$ solutions $\pi_i$ could
intersect. 

Let us analyse the intersections of the conformal
weights $\Delta_{r,s}(t)$ of the $\Pi_l$ solutions to 
$rs=2l$, $r,s\in\bbbn$, s odd.
For fixed $l\in\bbbnu$ we find the prime factorisation as 
given above. Let us now take two $P$-tuples
$\pi^1$ and $\pi^2$ ($\pi^1\not =\pi^2$) and let us assume that 
the corresponding conformal weights intersect:
$\Delta_{r_{\pi^1},s_{\pi^1}}= \Delta_{r_{\pi^2},s_{\pi^2}}$. 
Applying \eq{\ref{eq:deltars}} one gets straightforwardly that
this is the case if and only if
\bea
\left(2^n \prod_{i=1}^P p_i^{k_i^1} t + 
\prod_{i=1}^P p_i^{n_i-k_i^1}\right)^2 &=&
\left(2^n \prod_{i=1}^P p_i^{k_i^2} t + 
\prod_{i=1}^P p_i^{n_i-k_i^2}\right)^2 \pkt
\eea
This easily rearranges to two sign-symmetric solutions for the 
parameter $t$ of the central extension $c$:
\bea
t &=& \pm \frac{1}{2^n}\prod_{i=1}^P p_i^{n_i-k_i^1-k_i^2} \pkt 
\label{eq:intersect}
\eea
Again, let us consider the example of level $l=630$ and consider 
the two $3$-tuples $\pi^1=(1,1,1)$ and $\pi^2=(2,0,1)$ 
for the order $(3,5,7)$ of the relevant prime factors.
\eq{\ref{eq:intersect}} tells us that the conformal weights 
$\Delta_{r_{\pi^1},s_{\pi^1}}$ and
$\Delta_{r_{\pi^2},s_{\pi^2}}$ have exactly $2$ intersection 
points, namely $t=\pm \frac{1}{2^2\times 3\times 7}=\pm\frac{1}{84}$, 
where the two singular vectors $\Psi^+_{r_{\pi^1},s_{\pi^1}}$ and 
$\Psi^+_{r_{\pi^2},s_{\pi^2}}$ both exist. The key question will now be 
whether these singular vectors are actually different and hence
define a two-dimensional singular space, or if they are proportional 
and thus lead to a one-dimensional singular space instead.

For this purpose, let us again take two different $P$-tuples 
$\pi^1$ and $\pi^2$. The corresponding conformal weights intersect for 
$t=\pm \frac{1}{2^n}\prod_{i=1}^P p_i^{n_i-k_i^1-k_i^2}$. Using
conjecture \refoth{\ref{cj:ab}} we obtain the corresponding 
singular vectors
$\Psi^+_{r_{\pi^j},s_{\pi^j}}$, $j=1,2$ as:
\bea
\Psi^+_{r_{\pi^j},s_{\pi^j}} &=&
(r_{\pi^j}t+s_{\pi^j},4r_{\pi^j}t)^+_l\ket{\Delta_{r_{\pi^j},s_{\pi^j}}}
\com j=1,2 \pkt
\eea 
Inserting the expressions \eqs{\ref{eq:rspi}} for $r_{\pi^j}$ and 
$s_{\pi^j}$ and replacing
$t$ by \eq{\ref{eq:intersect}} we hence obtain 
\bea
\Psi^+_{r_{\pi^j},s_{\pi^j}} &=& \pm (\prod_{i=1}^P 
p_i^{n_i-k_i^{\bar{j}}} \pm \prod_{i=1}^P  p_i^{n_i-k_i^j} ,
 4 \prod_{i=1}^P  p_i^{n_i-k_i^{\bar{j}}}  )^+_l
\ket{\Delta_{r_{\pi^j},s_{\pi^j}}} \com j=1,2 \com
\eea 
with $\bar{1}=2$ and $\bar{2}=1$.
The determinant of the coefficients shows that 
$\Psi^+_{r_{\pi^1},s_{\pi^1}}$ and
$\Psi^+_{r_{\pi^2},s_{\pi^2}}$ are linearly dependent if and only if
\bea
\pm 4  \prod_{i=1}^P  [p_i^{n_i-k_i^1}]^2 \mp 4  \prod_{i=1}^P  
[p_i^{n_i-k_i^2}]^2 &=& 0 \com 
\eea
\bea 
\Leftrightarrow &&  \prod_{i=1}^P  p_i^{n_i-k_i^1} \;\; = \;\;  
\prod_{i=1}^P  p_i^{n_i-k_i^2} \com 
\eea
\bea
\Leftrightarrow && \pi^1 \;\; = \;\; \pi^2 \com    
\eea
in contradiction with our original assumption. Hence, the vectors 
$\Psi^+_{r_{\pi^1},s_{\pi^1}}$ and $\Psi^+_{r_{\pi^2},s_{\pi^2}}$ 
are linearly independent for $\pi^1\not =\pi^2$.
Similar arguments also show that $\Psi^-_{r_{\pi^1},s_{\pi^1}}$ and
$\Psi^-_{r_{\pi^2},s_{\pi^2}}$ are linearly independent
for $\pi^1\not =\pi^2$ (at exactly the same values for $t$).

An important observation now is the following. On the one hand,
we have proven that the maximal singular dimension is $2$ and, on the other, 
we have shown that for different $P$-tuples $\pi_1$ and $\pi_2$ the
corresponding singular vectors are always linearly independent at the 
intersection points (i.e. in the same Verma module). Therefore, since 
for high levels $l$ there are many different $P$-tuples, the question 
arises whether or not it is possible
that the conformal weights of more than two $P$-tuples can intersect 
at the same value $t$, obtaining a contradiction to either of our results. 
That this is not the case is shown in the following consideration. The 
conformal weights of $\pi_1$ and $\pi_2$ ($\pi_1\not = \pi_2$) intersect at
$t=\pm \frac{1}{2^n}\prod_{i=1}^P p_i^{n_i-k_i^1-k_i^2}$. In the same way 
the conformal weights of $\pi_2$ and $\pi_3$ ($\pi_2\not = \pi_3$) 
intersect at
$\bar{t}=\pm \frac{1}{2^n}\prod_{i=1}^P p_i^{n_i-k_i^2-k_i^3}$. 
Assuming $t=\bar{t}$ is equivalent to
\bea
\prod_{i=1}^P p_i^{n_i-k_i^1-k_i^2} &=& 
\prod_{i=1}^P p_i^{n_i-k_i^2-k_i^3} \com
\eea
and thus $\prod_{i=1}^P p_i^{k_i^1} = \prod_{i=1}^P p_i^{k_i^3}$ results
in $\pi_1=\pi_3$ due to the fact that the numbers $p_i$ are distinct primes. 
Therefore, there are no values of $t$ for which more than two conformal 
weights $\Delta_{r_{\pi},s_{\pi}}$ intersect for different 
$P$-tuples $\pi$. We should stress again, that these results are based 
on the conjecture \refoth{\ref{cj:ab}} which seems to be a very 
reliable conjecture. We summarise our results in the following theorem.

\bth \label{th:prsvecs}
Let us fix a level $l\in\bbbnu$. Based on the conjecture 
\refoth{\ref{cj:ab}} one finds that at the level 
$l=\frac{rs}{2}$ the singular vectors $\Psi^{\pm}_{r,s}$ 
define generically one-dimensional singular spaces. The only 
exceptions, where degenerate singular spaces of dimension $2$
arise, occur necessarily for the cases when two conformal weights 
$\Delta_{r_1,s_1}(t)$ and $\Delta_{r_2,s_2}(t)$
intersect. For each two different pairs $(r_1,s_1)$ 
and $(r_2,s_2)$ with $l=\frac{r_1s_1}{2}=\frac{r_2s_2}{2}$,
$r_1,r_2,s_1,s_2\in\bbbn$, $s_1,s_2$ odd, there are 
two values of $t$ where such an intersection happens. These 
values are $t=\pm \frac{1}{2^n}\prod_{i=1}^P p_i^{n_i-k_i^1-k_i^2}$ and  
they are real and rational in all cases.
However, the conformal weights of three such pairs do 
not intersect for any common values of $t$. 
\eth

%%%%%%%%%%%%%%%%%%%%%%%%%%%%%%%%%%%%%%%%%%%%%%%%%%%%%%%%%%%
%                                                         %
%                G-closed Classification                  %
%                                                         %
%%%%%%%%%%%%%%%%%%%%%%%%%%%%%%%%%%%%%%%%%%%%%%%%%%%%%%%%%%%

\section{$G$-closed Verma modules.}
\label{sec:GclosedVM}

We will now consider $G$-closed Verma modules $\vm_{\frac{c}{24}}^G$ 
and discuss the relation between the singular vectors in these 
(incomplete) Verma modules and the singular vectors in complete 
Verma modules $\vm_{\Delta}$ with $\Delta=\frac{c}{24}$.
For this value the Verma module $\vm_{\frac{c}{24}}$ contains 
a singular vector $\Psi^-_{0} = G_0 \ket{\frac{c}{24}}$ at level 0 
with parity $-$. In order to
obtain irreducible representations one considers the quotient space 
\bea
\vm_{\frac{c}{24}}^G &=& \frac{\vm_{\frac{c}{24}}}{U(\tw)\Psi^-_{0}} \com
\eea
which is precisely the $G$-closed Verma module $\vm_{\frac{c}{24}}^G$. 
In fact, this quotient module is only one among the quotient modules. 
We could equally well consider quotients
with respect to modules generated by singular vectors $\Psi_{r,s}^{\pm}$ 
at levels $\frac{rs}{2}$. 
The reason why we are especially interested in $\vm_{\frac{c}{24}}^G$
is twofold. First, for physical applications it could be relevant
if a highest weight vector (or a singular vector) is $G$-closed,
which makes $\vm_{\frac{c}{24}}^G$ special compared to other quotient
spaces. Secondly, the Verma module $\vm_{\frac{c}{24}}$
is easier to construct by explicit calculations than most Verma modules 
with singular vectors. Nevertheless, we aim also at
obtaining information on the structure of quotient modules that is valid
not only for $\vm_{\frac{c}{24}}^G$ but also for quotient modules 
of similar type.

Via the canonical map defined for quotient spaces, singular vectors from
$\vm_{\frac{c}{24}}$ are either also singular in $\vm_{\frac{c}{24}}^G$ or
they are trivial (they `go away'). The latter happens if and only if the 
singular vector in question is
a descendant vector of the singular vector $\Psi_{0}^-$ in 
$\vm_{\frac{c}{24}}$.
Thus, singular vectors of $\vm_{\frac{c}{24}}^G$ can either be 
inherited from
$\vm_{\frac{c}{24}}$ via the canonical quotient map or they 
appear for the first time in the quotient
$\vm_{\frac{c}{24}}^G$ as vectors that were not singular in 
$\vm_{\frac{c}{24}}$.
In the latter case these singular vectors of 
$\vm_{\frac{c}{24}}^G$ are called subsingular vectors
in $\vm_{\frac{c}{24}}$. The Virasoro algebra does not allow 
any subsingular vectors
which is a consequence of the proof of Feigin and Fuchs\cite{ff1}, 
neither do
the $N=1$ superconformal algebras\cite{ast1}. Surprisingly enough, the
$N=2$ superconformal algebras contain subsingular vectors which have
first been discovered
in Refs. \icite{beatriz1,beatriz2} for the Neveu-Schwarz, the
Ramond, and the topological $N=2$ algebras. So far, 
nothing was known about the existence
of subsingular vectors in the twisted $N=2$ case. Due to the very 
different structure of this particular
$N=2$ algebra it was hard to say what one {\it a priori} should expect.
However, we will show 
in this section that the twisted $N=2$ superconformal algebra also contains
subsingular vectors.

In section \refoth{\ref{sec:sdim}} we have found that the singular 
dimensions for $G$-closed
Verma modules are just $1$. Therefore, there are no degenerate 
singular vectors in $G$-closed Verma modules. A singular vector 
in $\vm_{\frac{c}{24}}^G$ at level $l$ can therefore be identified by 
the coefficient of only one particular term (being zero or non-zero). 
The ordering kernel in \tab{\ref{tab:adkern3}} tells us that this 
particular term is $T_{-\half}^{2l}$ for the positive parity case and
$T_{-\half}^{2l-1}G_{-\half}$
for the negative parity one. We can thus give the following theorem.
\bth \label{th:Gvanish}
Two singular vectors in the $G$-closed Verma module 
$\vm_{\frac{c}{24}}^G$ at the same 
level and with the same parity are always proportional. 
If a vector satisfies the highest weight
conditions at level $l$, with parity $+$ or parity $-$, 
and its coefficient for the term
$T_{-\half}^{2l}$ or $T_{-\half}^{2l-1}G_{-\half}$ 
vanishes, respectively, then this vector is trivial.
\eth
   
Before looking at explicit examples of singular vectors in 
$\vm_{\frac{c}{24}}^G$ we should first analyse whether or not 
the existing singular vectors $\Psi_{r,s}^{\pm}$ 
in $\vm_{\frac{c}{24}}$ are descendants of $\Psi_{0}^-$. 
Requiring that $\Delta_{r,s}(t^G)=\frac{c}{24}$ 
in \eq{\ref{eq:deltars}} one identifies easily
the values $t^G$ of $t$ for which $\Psi_{r,s}^{\pm}$ exists 
in $\vm_{\frac{c}{24}}$ at level $\frac{rs}{2}$:
\bea
t^G_{r,s} &=& -\frac{s}{r} \com
\eea
for $r,s\in\bbbn$, $s$ odd. We observe that if 
$\Psi_{r,s}^{\pm}\in\vm_{\frac{c}{24}}$, which happens
only for the values $t^G_{r,s}$, then one finds also the 
singular vectors $\Psi_{nr,ns}^{\pm}\in \vm_{\frac{c}{24}}$ 
at levels $\frac{n^2rs}{2}$ for any positive integer $n\in\bbbn$. 
Conjecture \refoth{\ref{cj:ab}}
gives us the explicit form of the vectors $\Psi_{r,s}^{\pm}$. From
$\Psi_{r,s}^+=(rt+s,4rt)^+_{\frac{rs}{2}}\ket{\frac{c}{24}}$ 
one obtains for the values of $t^G_{r,s}$ that
$\Psi_{r,s}^+(t^G_{r,s})=(0,-4s)^+_{\frac{rs}{2}}\ket{\frac{r-s}{8r}}$,
and similarly
$\Psi_{r,s}^-=(-\frac{2}{r},rt+s)^-_{\frac{rs}{2}}\ket{\frac{c}{24}}$
leads to $\Psi_{r,s}^-(t^G_{r,s})=
(-\frac{2}{r},0)^-_{\frac{rs}{2}}\ket{\frac{r-s}{8r}}$.
As the coefficient of $T_{-\half}^{rs}$ for the positive parity case
and the coefficient of $T_{-\half}^{rs-1}G_{-\half}$ 
in the negative parity case vanish, due to theorem \refoth{\ref{th:Gvanish}}
we hence find that the singular vectors $\Psi_{r,s}^{\pm}(t^G_{r,s})$ vanish
in the quotient space $\vm_{\frac{c}{24}}^G$. Therefore the singular 
vectors $\Psi_{r,s}^{\pm}(t^G_{r,s})$ are descendant
vectors of $\Psi_{0}^-(t^G_{r,s})$. 
Consequently\footnote{Strictly speaking, we have to assume that there are
no isolated singular vectors for the twisted $N=2$ superconformal 
algebra. We have not been able to compute
any isolated singular vectors neither have they ever 
appeared for any other chiral algebra.}
all the singular vectors that can be found in $\vm_{\frac{c}{24}}^G$
are subsingular in $\vm_{\frac{c}{24}}$.

By explicit computation one easily finds that singular vectors 
$\Upsilon_{r,s}^{\pm}$ do appear in $\vm_{\frac{r-s}{8r}}^G$ at
exactly the levels $\frac{rs}{2}$ where the singular vectors 
$\Psi_{r,s}^{\pm}$ disappear in the quotient kernel. (The reasons for 
this will will be shown in a future publication\cite{p11vanish}).
In addition these are the only singular vectors
appearing in $\vm_{\frac{c}{24}}^G$.
The vectors $\Upsilon_{r,s}^{\pm}$ are hence subsingular vectors in 
the complete Verma modules
$\vm_{\frac{r-s}{8r}}$. This has been verified for levels $\half$,
$1$, and $\thalf$. The explicit results are given in 
appendix \refoth{\ref{app:b}}.

We conclude this section with another interesting fact about the 
$G$-closed Verma modules.
As we know, the singular dimension for $G$-closed Verma modules is 
only $1$. One
may wonder what happens in the cases where $\vm_{\frac{c}{24}}$ has 
a two-dimensional singular space. First of all, we have just shown 
that if this happens then - due to the nature of the singular vectors 
$\Psi_{r,s}^{\pm}$ - the whole two-dimensional singular space
would vanish in the quotient space $\vm_{\frac{c}{24}}^G$ as it is
descendant of $\Psi_{0}^-$.
Nevertheless, one could argue that for these cases the corresponding 
subsingular vectors that appear in a one to one correspondence at the 
same levels than the singular vectors that disappear, may 
span a two-dimensional subsingular space. However, Verma modules
$\vm_{\Delta}$ that contain degenerate singular vectors never satisfy
${\Delta=\frac{c}{24}}$ and therefore this question is redundant. Namely,
if we use in $t=-\frac{s}{r}$ the expressions of \eqs{\ref{eq:rspi}} 
for $r$ and $s$ 
and require that $t=\pm \frac{1}{2^n}\prod_{i=1}^P p_i^{n_i-k_i^1-k_i^2}$ 
in the case of degeneracy for
two $P$-tuples $\pi^1$ and $\pi^2$, one easily finds 
\bea
\pm \frac{1}{2^n}\prod_{i=1}^P p_i^{n_i-k_i^1-k_i^2} &=& 
- \frac{1}{2^n}\prod_{i=1}^P p_i^{n_i-2k_i^1} \pkt \label{eq:doubleG}
\eea
Taking into account that the $p_i$ are distinct primes then the 
only solution to \eq{\ref{eq:doubleG}} is, however, $\pi^1=\pi^2$, 
which shows that the modules $\vm_{\Delta}$ with $\Delta=\frac{c}{24}$
never have two-dimensional singular spaces.

%%%%%%%%%%%%%%%%%%%%%%%%%%%%%%%%%%%%%%%%%%%%%%%%%%%%%%%%%%%
%                                                         %
%                G-closed singular vectors                %
%                                                         %
%%%%%%%%%%%%%%%%%%%%%%%%%%%%%%%%%%%%%%%%%%%%%%%%%%%%%%%%%%%

\section{$G$-closed singular vectors.}
\label{sec:Gclosedsvecs}

In this last section we will analyse $G$-closed singular vectors in 
a very similar way
as we have analysed $G$-closed Verma modules in the previous section. 
Due to the constraints 
on the conformal weights, there are no $G$-closed singular vectors 
in $G$-closed Verma modules,
as mentioned already earlier. Therefore, we take 
$\Psi_{r,s}^{\pm}(t)\in\vm_{\Delta_{r,s}(t)}$ and impose the
conformal weight constraint $\Delta_{r,s}+\frac{rs}{2}=\frac{c}{24}$ 
which is satisfied by $G$-closed singular vectors. The condition 
$\Delta_{r,s}+\frac{rs}{2}=\frac{c}{24}$ has the unique solution
\bea
{}^Gt_{r,s} &=& \frac{s}{r} \pkt
\eea
Let us note the similarity between ${}^Gt_{r,s}$ and $t^G_{r,s}$ of 
the previous section.

According to the result that the singular dimension for $G$-closed 
singular vectors is $1$, the ordering kernel being given in table
\tab{\ref{tab:adkern2}}, a $G$-closed singular vector $\Psi_{l}^{\pm G}$ 
at level $l=\frac{c}{24}-\Delta$ can be identified  
by the coefficient of $T_{-\half}^{2l}$ or $T_{-\half}^{2l}G_0$
(being zero or non-zero), depending on the parity.
We therefore give the following theorem.
\bth \label{th:Gsvec}
Two $G$-closed singular vectors in the same Verma module $\vm_{\Delta}$,
at the same level $l=\frac{c}{24}-\Delta$  
and with the same parity are always proportional. 
If a $G$-closed vector satisfies the highest weight
conditions at level $l=\frac{c}{24}-\Delta$ and
parity $+$, or parity $-$, and its coefficient for the term
$T_{-\half}^{2l}$ or $T_{-\half}^{2l}G_{0}$ vanishes, 
respectively, then this vector is trivial.
\eth

Using the expressions for  ${}_G(a,b)^{\mp}_l$ given by
\eqs{\ref{eq:thetaG1}}-\eqoth{\ref{eq:thetaG4}}, we obtain 
for the vector $G_0\Psi_{r,s}^{+}
={}_G(rt+s,4rt)^-_{\frac{rs}{2}}\ket{\Delta_{r,s}}$, in the case  
$t={}^Gt_{r,s}$, the result $G_0\Psi_{r,s}^{+}
=(0,0)^-_{\frac{rs}{2}}\ket{\Delta_{r,s}}$. Since this vector 
satisfies the highest weight conditions but has 
vanishing coefficient for the term $T_{-\half}^{rs} G_0$, we find that 
$G_0\Psi_{r,s}^{+}\equiv 0$. Thus the singular vectors $\Psi_{r,s}^{+}$
are $G$-closed for $t={}^Gt_{r,s}=\frac{s}{r}$. Similarly, we obtain 
for  $G_0\Psi_{r,s}^{-}
={}_G(-\frac{2}{r},rt+s)^+_{\frac{rs}{2}}\ket{\Delta_{r,s}}$, in the case  
$t={}^Gt_{r,s}$, $G_0\Psi_{r,s}^{-}
=(0,0)^+_{\frac{rs}{2}}\ket{\Delta_{r,s}}$ and hence the singular
vectors $\Psi_{r,s}^{-}$ are
also $G$-closed for $t={}^Gt_{r,s}=\frac{s}{r}$. It is surprising the
fact that for $t={}^Gt_{r,s}$ all the singular vectors become $G$-closed
(this is not required by first principles).
\bth
The singular vectors $\Psi_{r,s}^{\pm}\in\vm_{\Delta_{r,s}}$ 
become $G$-closed exactly for the 
values of the central parameter $t={}^Gt_{r,s}=\frac{s}{r}$, 
$r,s\in\bbbn$, $s$ odd.
\eth

Like in the previous section, let us finally check that the 
cases where $\Psi_{r,s}^{\pm}\in\vm_{\Delta_{r,s}}$ 
become $G$-closed cannot be degenerate cases, as suggested by the 
results of table \tab{\ref{tab:adkern2}}.
We set $t=\frac{s}{r}$ and use again the expressions of 
\eqs{\ref{eq:rspi}} for $r$ and $s$. 
Requiring that $t=\pm \frac{1}{2^n}\prod_{i=1}^P 
p_i^{n_i-k_i^1-k_i^2}$ in the case of degeneracy for
two different $P$-tuples $\pi^1$ and $\pi^2$ we obtain 
\bea
\pm \frac{1}{2^n}\prod_{i=1}^P p_i^{n_i-k_i^1-k_i^2} &=& 
+ \frac{1}{2^n}\prod_{i=1}^P p_i^{n_i-2k_i^1} \pkt \label{eq:doubleGsvec}
\eea
Again, the only solution to \eq{\ref{eq:doubleGsvec}} is
$\pi^1=\pi^2$, in contradiction with the assumption that 
$\pi^1$ and $\pi^2$ are distinct. This shows that the degenerate cases do
not correspond to $G$-closed singular vectors,
i.e. two-dimensional singular spaces are never $G$-closed.

%%%%%%%%%%%%%%%%%%%%%%%%%%%%%%%%%%%%%%%%%%%%%%%%%%%%%%%%%%%
%                                                         %
%                Conclusions and prospects                %
%                                                         %
%%%%%%%%%%%%%%%%%%%%%%%%%%%%%%%%%%%%%%%%%%%%%%%%%%%%%%%%%%%

\section{Conclusions and prospects}
\label{sec:conclusions}

In this paper we have applied the method of adapted orderings to the 
twisted $N=2$ superconformal algebra. This shows, again, that this 
method is a powerful tool that can easily be applied to a large 
classes of algebras. The way it works is rather simple and it 
is very flexible in its use. Whereas the application of this method to
the topological, Neveu-Schwarz and Ramond $N=2$ algebras
in Ref. \icite{p6sdim1} assigned a special
r\^ole to the Virasoro generator $L_{-1}$, for the twisted $N=2$
algebra this special r\^ole has been taken over by the U(1) current
generator $T_{-\half}$. Another remarkable difference compared to our
earlier applications of this method is that in the twisted $N=2$ case
the fermionic field modes need to be mixed in the adapted ordering
rather than being kept separate.

As a result we have derived the singular dimensions and the ordering 
kernels for the twisted $N=2$ algebra. This reveals that also the
twisted $N=2$ algebra contains degenerate singular
spaces with dimension $2$ in complete Verma modules, as it is the case 
for the three isomorphic $N=2$ algebras.
On the other hand, $G$-closed Verma modules have singular dimension
$1$ and therefore do not allow any degenerate singular spaces.

Based on the ordering kernel coefficients we have conjectured general
expressions for the relevant terms of all (primitive) singular vectors. 
These expressions lead to the identification of all degenerate 
cases with singular dimension 2 
and also to the identification of all $G$-closed singular vectors.
These expressions also lead to the discovery of subsingular vectors 
for the twisted $N=2$ 
superconformal algebra. As in many examples for the other $N=2$
superconformal algebras, these subsingular vectors appear in relation
to additional vanishing conditions on a singular vector. This relation
can be generally proven and will be given in a forthcoming 
paper\cite{p11vanish}. We have seen that ordering kernel coefficients
play a crucial r\^ole in this matter.

We have also computed multiplication rules 
for singular vector operators, which sets the foundation for the study
of the twisted $N=2$ embedding diagrams which will be investigated in a
future publication.

The twisted $N=2$ superconformal algebra has hence proven to have a
rather rich structure that contains many features that were also 
discovered for the three isomorphic $N=2$ superconformal
algebras. Nevertheless, the way these features appear in the case 
of the twisted $N=2$ algebra is quite different from the way they appear 
for the other $N=2$ algebras. For example, the degenerate singular
spaces are produced simply by the intersection of two 
primitive singular vectors. For no other $N=2$ algebra primitive 
singular vectors at the same level and with the same charge are
linearly independent. Furthermore, the discovery of degenerate
singular vectors in the twisted $N=2$ case is rather surprising as the
underlying parameter space is just one-dimensional. The
twisted $N=2$ superconformal algebra hence turns out to be 
quite distinct from all other superconformal algebras analysed so
far in the literature. This raises the need for the analysis of
twisted superconformal algebras for higher $N$ - a very
interesting question already because the periodicity
conditions of more than 2 fermionic fields can be mixed in different
ways. The concept of adapted orderings should apply also in these cases. 

%%%%%%%%%%%%%%%%%%%%%%%%%%%%%%%%%%%%%%%%%%%%%%%%%%%%%%%%%%%
%                                                         %
%                Appendix A                               %
%                                                         %
%%%%%%%%%%%%%%%%%%%%%%%%%%%%%%%%%%%%%%%%%%%%%%%%%%%%%%%%%%%

\appendix
\section{Examples of negative parity singular vectors}
\label{app:a}
For completeness we shall also give all singular vectors with 
negative parity
at level $\frac{1}{2}$, $1$, and $\frac{3}{2}$. They were found 
by explicit computer calculations\cite{maple}.

\noi 
Level $\frac{1}{2}$:
\bea
\Psi^-_{1,1}(t) &=& \Bigl\{ -2 T_{-\half}G_0 +(t+1) G_{-\half} \Bigr\} 
\ket{-\frac{t+1}{8t}} \pkt
\eea 

\noi
Level $1$:
\bea
\Psi^-_{2,1}(t) &=& \Bigl\{ - T_{-\half}^2G_0 +(2t+1) T_{-\half} G_{-\half} 
-2t L_{-1}G_0 \nn \\
&& -\frac{t}{2}(2t+3) G_{-1} \Bigr\}
\ket{-\frac{3t^2+3t+1}{8t}} \pkt
\eea 

\noi
Level $\frac{3}{2}$ has two different singular vectors:
\bea
\Psi^-_{3,1}(t) &=& \Bigl\{  -\frac{2}{3} T_{-\half}^3 G_0 + (3t+1)
T_{-\half}^2G_{-\half} 
-4tL_{-1}T_{-\half}G_0 \nn \\
&&   -\frac{1}{3}(12t^2+13t+3) G_{-1}T_{-\half} + \frac{2t}{3}(3t+1)
L_{-1}G_{-\half} 
-\frac{4t}{3} G_{-1}G_{-\half}G_0  \nn\\
&& -\frac{4t}{3}(t-1) T_{-\thalf}G_0 +
(2t^3-\frac{4}{3}t^2-\frac{8}{3}t-\frac{2}{3})  
G_{-\thalf} \Bigr\} \ket{-\frac{8t^2+5t+1}{8t}} \com \\
\Psi^-_{1,3}(t) &=&\Bigl\{  -2T_{-\half}^3  G_0+(t+3)
T_{-\half}^2G_{-\half} 
+4tL_{-1}T_{-\half}G_0 \nn \\
&&   -(t+3)G_{-1}T_{-\half} - 2t(t+3) L_{-1}G_{-\half} 
-4t G_{-1}G_{-\half} G_0 \nn\\
&& +8 T_{-\thalf}G_0 -2(t+3) G_{-\thalf} \Bigr\} 
\ket{-\frac{5t+9}{8t}} \pkt
\eea 

In the notation of definition \refoth{\ref{def:ab}} we obtain:
\bea
\Psi^-_{1,1}(t) &=& (-2,t+1)^-_{\half} \ket{-\frac{t+1}{8t}}  \com \\
\Psi^-_{2,1}(t) &=& (-1,2t+1)^-_{1} \ket{-\frac{3t^2+3t+1}{8t}}  \com \\
\Psi^-_{3,1}(t) &=& (-\frac{2}{3},3t+1)^-_{\thalf} 
\ket{-\frac{8t^2+5t+1}{8t}}  \com \\
\Psi^-_{1,3}(t) &=& (-2,t+3)^-_{\thalf} \ket{-\frac{5t+9}{8t}}  \pkt
\eea

For $t=\pm 1$ we observe that $\Psi^-_{3,1}$ and
$\Psi^-_{1,3}$ span a two-dimensional singular vector space,
as it was the case for their positive parity partners.
\bea
\Psi^-_{3,1}(1)&=& (-\frac{2}{3},4)^-_{\thalf} \ket{-\frac{7}{4}}  \com \\
\Psi^-_{1,3}(1) &=& (-2,4)^-_{\thalf} \ket{-\frac{7}{4}}  \com \\
\Psi^-_{3,1}(-1)&=& (-\frac{2}{3},-2)^-_{\thalf} \ket{\frac{1}{2}}  \com \\
\Psi^-_{1,3}(-1) &=& (-2,2)^-_{\thalf} \ket{-\frac{1}{2}}  \pkt
\eea

%%%%%%%%%%%%%%%%%%%%%%%%%%%%%%%%%%%%%%%%%%%%%%%%%%%%%%%%%%%
%                                                         %
%                Appendix B                               %
%                                                         %
%%%%%%%%%%%%%%%%%%%%%%%%%%%%%%%%%%%%%%%%%%%%%%%%%%%%%%%%%%%

\section{Examples of singular vectors in $G$-closed Verma modules}
\label{app:b}

As discussed in section \refoth{\ref{sec:GclosedVM}}, the singular 
vectors $\Psi_{r,s}^{\pm}\in\vm_{\Delta_{r,s}}$, at levels $\frac{rs}{2}$,
are always descendants of the level zero singular vector 
$\Psi_0^- = G_0\ket{\frac{c}{24}}$ in the cases where 
$\Delta_{r,s}(t)=\frac{c}{24}$. This happens for the values 
$t=t^G_{r,s}=-\frac{s}{r}$, so that $\frac{c}{24}=\frac{r-s}{8r}$. 
Therefore, the singular vectors $\Psi_{r,s}^{\pm}$ are absent
in the $G$-closed Verma modules 
$\vm_{\frac{c}{24}}^G = \frac{\vm_{\frac{c}{24}}}{G_0\ket{\frac{c}{24}}}$.
All singular vectors which we find in $\vm^G_{\frac{c}{24}}$
are therefore subsingular vectors in
$\vm_{\frac{c}{24}}$. Surprisingly enough, the only 
singular vectors we have found in $\vm^G_{\frac{c}{24}}$ are
exactly at the levels and for the modules where we would have expected 
$\Psi_{r,s}^{\pm}$ to appear, if they were not trivial in the quotient
module. That is, they are located at levels $\frac{rs}{2}$ in the
Verma modules with $t=-\frac{s}{r}$.
In the following we will give all the singular vectors
in $\vm^G_{\frac{c}{24}}$ for levels $\half$, $1$, and $\thalf$.
These vectors are therefore subsingular vectors in 
$\vm_{\frac{c}{24}}$.

\noi
Level $\half$ has singular vectors only for $t=-1$:
\bea
\Upsilon_{1,1}^{+}(-1) &=& T_{-\half}\ket{0}^G \com \\
\Upsilon_{1,1}^{-}(-1) &=& G_{-\half}\ket{0}^G \pkt
\eea

\noi
Level $1$ has singular vectors only for $t=-\half$:
\bea
\Upsilon_{2,1}^{+}(-\half) &=&  \Bigl\{ T_{-\half}^2 -L_{-1} \Bigr\} 
\ket{\frac{1}{16}}^G \com \\
\Upsilon_{2,1}^{-}(-\half) &=& \Bigl\{ 4T_{-\half} G_{-\half} -G_{-1} 
\Bigr\} 
\ket{\frac{1}{16}}^G \pkt
\eea

\noi
Level $\thalf$ has singular vectors only for $t=-\frac{1}{3}$ and for
$t=-3$:
\bea
\Upsilon_{3,1}^{+}(-\frac{1}{3}) &=& \Bigl\{ 9T_{-\half}^3 -18
L_{-1}T_{-\half} -6 G_{-1}G_{-\half} +8 T_{-\thalf} \Bigr\}
\ket{\frac{1}{12}}^G \com \\
\Upsilon_{3,1}^{-}(-\frac{1}{3}) &=& \Bigl\{ 27T_{-\half}^2G_{-\half}
-15G_{-1}T_{-\half}-6L_{-1}G_{-\half}-10G_{-\thalf} \Bigr\}
\ket{\frac{1}{12}}^G \pkt
\eea
\bea
\Upsilon_{1,3}^{+}(-3) &=& \Bigl\{ T_{-\half}^3 +6
L_{-1}T_{-\half} -6 G_{-1}G_{-\half} -4 T_{-\thalf} \Bigr\}
\ket{-\frac{1}{4}}^G \com \\
\Upsilon_{1,3}^{-}(-3) &=& \Bigl\{ T_{-\half}^2G_{-\half}
-G_{-1}T_{-\half}+6L_{-1}G_{-\half}-2G_{-\thalf} \Bigr\}
\ket{-\frac{1}{4}}^G \pkt
\eea
If we replace $\ket{\frac{r-s}{8r}}^G$ by $\ket{\frac{r-s}{8r}}$ 
the resulting singular vectors are thus subsingular vectors in the
Verma modules $\vm_{\frac{r-s}{8r}}$.

The appearance of these subsingular vectors is very closely related
with an additional vanishing condition for the singular vector
$G_0\ket{\frac{c}{24}}$. That such an additional vanishing condition
results in an additional null vector which is most likely subsingular
will be proven in a forthcoming publication\cite{p11vanish}.

%%%%%%%%%%%%%%%%%%%%%%%%%%%%%%%%%%%%%%%%%%%%%%%%%%%%%%%%%%%
%                                                         %
%                Acknowledgements                         %
%                                                         %
%%%%%%%%%%%%%%%%%%%%%%%%%%%%%%%%%%%%%%%%%%%%%%%%%%%%%%%%%%%

\acknowledgements
We are grateful to Adrian Kent for many discussions on the ordering method 
and to Victor Kac for his explanations on superconformal algebras. M.D. is
indebted to the Deutsche Forschungsgemeinschaft (DFG) for financial
support and to DAMTP for the hospitality.

\noi

%%%%%%%%%%%%%%%%%%%%%%%%%%%%%%%%%%%%%%%%%%%%%%%%%%%%%%%%%%%
%                                                         %
%                Bibliography                             %
%                                                         %
%%%%%%%%%%%%%%%%%%%%%%%%%%%%%%%%%%%%%%%%%%%%%%%%%%%%%%%%%%%

%\bibliography{../ref}
%\bibliographystyle{plain}

\end{document}